\documentclass[12pt,a4paper]{article}
\usepackage{feynmp,amsmath,amssymb}
\usepackage{jheppub}
\usepackage{hyperref}
\newcommand{\LL}{\mathcal{L}}
\newcommand{\pd}{\partial}
\newcommand{\tr}{\mathrm{Tr}\,}
\newcommand{\um}{\mathbf{1}}

\newcommand{\ii}{\mathrm{i}}

\makeatletter
\font\manfnt=manfnt
\def\Watchout{\@ifnextchar [{\W@tchout}{\W@tchout[1]}}
\def\W@tchout[#1]{{\manfnt\@tempcnta#1\relax%
  \@whilenum\@tempcnta>\z@\do{%
    \char"7F\hskip 0.3em\advance\@tempcnta\m@ne}}}
\def\dubious{\@ifnextchar[{\@dubious}{\@dubious[1]}}

\def\@dubious[#1]{%
  \setbox\@tempboxa\hbox{\@W@tchout#1}
  \@tempdima\wd\@tempboxa
  \list{}{\leftmargin\@tempdima}\item[\hbox to 0pt{\hss\@W@tchout#1}]}
\def\thomega{\@ifnextchar[{\@thomega}{\@thomega[1]}}

\def\@thomega[#1]{%
  \setbox\@tempboxa\hbox{\@W@tchout#1}
  \@tempdima\wd\@tempboxa
  \list{}{\leftmargin\@tempdima}\item[\hbox to 0pt{\hss\@W@tchout#1}]%
    \footnotesize
    \itshape\textbf{Treasures from Thorsten's little chest of formulae:}\quad}
\def\@W@tchout#1{\W@tchout[#1]}
\newif\if@preliminary
\@preliminaryfalse
\def\preliminary{\@preliminarytrue}
\def\fmslash{\@ifnextchar[{\fmsl@sh}{\fmsl@sh[0mu]}}
\def\fmsl@sh[#1]#2{%
  \mathchoice
    {\@fmsl@sh\displaystyle{#1}{#2}}%
    {\@fmsl@sh\textstyle{#1}{#2}}%
    {\@fmsl@sh\scriptstyle{#1}{#2}}%
    {\@fmsl@sh\scriptscriptstyle{#1}{#2}}}
\def\@fmsl@sh#1#2#3{\m@th\ooalign{$\hfil#1\mkern#2/\hfil$\crcr$#1#3$}}
\makeatother

\usepackage{graphicx}
\usepackage{feynmp}
\usepackage{amsmath}
\allowdisplaybreaks
\begin{document}

\begin{fmffile}{\jobname pics}
\setlength{\unitlength}{1mm}
\fmfset{arrow_ang}{10}
\fmfset{curly_len}{2mm}
\fmfset{wiggly_len}{3mm}
\newcommand{\setupFourAmp}{%
    \fmfleft{i1,i2}
    \fmfright{o1,o2}
    \fmf{phantom}{i1,v1,i2}
    \fmf{phantom}{o2,v2,o1}
    \fmf{phantom}{v1,v2}
    \fmffreeze
}
\newcommand{\setupSquare}{%
  \fmfleft{x1,i1,x2,x3,x4,i2,x5}
  \fmfright{y1,o1,y2,y3,y4,o2,y5}
  \fmftop{t}
  \fmfbottom{b}
  \fmf{phantom}{i1,c1,o1}
  \fmf{phantom}{o2,c2,i2}
  \fmf{dashes}{t,b}
  \fmffreeze
  \fmf{phantom}{i1,v1,c1,w1,o1}
  \fmf{phantom}{o2,w2,c2,v2,i2}
  \fmf{phantom}{v1,v2}
  \fmf{phantom}{w1,w2}
  \fmffreeze
}
\newcommand{\setupqqggAmp}{%
  \fmfipath{loop}
  \fmfiset{loop}{halfcircle rotated 90 scaled 2w shifted (w,h/2)}
  \fmfipath{prqt,prqu,prql,prqb}
  \fmfiset{prqt}{subpath ((0/4,1/4)*length loop) of loop}
  \fmfiset{prqu}{subpath ((1/4,2/4)*length loop) of loop}
  \fmfiset{prql}{subpath ((2/4,3/4)*length loop) of loop}
  \fmfiset{prqb}{subpath ((3/4,4/4)*length loop) of loop}
  \fmfipair{vqt,vqb,vgu,vgl,vgc,vqu,vql,vqc}
  \fmfiset{vqt}{point (0/4 * length loop) of loop}
  \fmfiset{vqu}{point (1/4 * length loop) of loop}
  \fmfiset{vqc}{point (2/4 * length loop) of loop}
  \fmfiset{vql}{point (3/4 * length loop) of loop}
  \fmfiset{vqb}{point (4/4 * length loop) of loop}
  \fmfiset{vgu}{(xpart vqt, ypart vqc + 1/3 ypart (vqt - vqc))}
  \fmfiset{vgl}{(xpart vqt, ypart vqc - 1/3 ypart (vqt - vqc))}
  \fmfipath{gloop}
  \fmfiset{gloop}{halfcircle rotated 90 scaled (ypart vgu - ypart vgl)
                  shifted (w, h/2)}
  \fmfiset{vgc}{point (1/2 * length gloop) of gloop}
  \fmfipath{prgc,prgu,prgl,prgt,prgb,prgf,prgr}
  \fmfiset{prgc}{vqc -- vgc}
  \fmfiset{prgu}{subpath ((0, 1/2) * length gloop) of gloop}
  \fmfiset{prgl}{subpath ((1/2, 1) * length gloop) of gloop}
  \fmfiset{prgt}{vqu .. {dir 0} vgu}
  \fmfiset{prgb}{vql .. {dir 0} vgl}
  \fmfiset{prgf}{vqu .. {dir 0} vgl}
  \fmfiset{prgr}{vql .. {dir 0} vgu}
}
\newcommand{\setupqqgg}{%
  \fmfi{dashes}{(0.5w,0h)--(0.5w,1h)}
  \fmfipath{loop}
  \fmfiset{loop}{fullcircle rotated 90 scaled 0.8w shifted (w/2,h/2)}
  \fmfipath{prqtl,prqul,prqll,prqbl}
  \fmfiset{prqtl}{subpath ((0/8,1/8)*length loop) of loop}
  \fmfiset{prqul}{subpath ((1/8,2/8)*length loop) of loop}
  \fmfiset{prqll}{subpath ((2/8,3/8)*length loop) of loop}
  \fmfiset{prqbl}{subpath ((3/8,4/8)*length loop) of loop}
  \fmfipath{prqtr,prqur,prqlr,prqbr}
  \fmfiset{prqtr}{subpath ((7/8,8/8)*length loop) of loop}
  \fmfiset{prqur}{subpath ((6/8,7/8)*length loop) of loop}
  \fmfiset{prqlr}{subpath ((5/8,6/8)*length loop) of loop}
  \fmfiset{prqbr}{subpath ((4/8,5/8)*length loop) of loop}
  \fmfipair{vqt,vqb,vgu,vgl,vgcl,vgcr,vqul,vqll,vqur,vqlr,vqcl,vqcr}
  \fmfiset{vqt }{point (0/8 * length loop) of loop}
  \fmfiset{vqul}{point (1/8 * length loop) of loop}
  \fmfiset{vqcl}{point (2/8 * length loop) of loop}
  \fmfiset{vqll}{point (3/8 * length loop) of loop}
  \fmfiset{vqb} {point (4/8 * length loop) of loop}
  \fmfiset{vqur}{point (7/8 * length loop) of loop}
  \fmfiset{vqcr}{point (6/8 * length loop) of loop}
  \fmfiset{vqlr}{point (5/8 * length loop) of loop}
  \fmfiset{vgu}{(xpart vqt, ypart vqcl + 1/3 ypart (vqt - vqcl))}
  \fmfiset{vgl}{(xpart vqt, ypart vqcl - 1/3 ypart (vqt - vqcl))}
  \fmfipath{gloop}
  \fmfiset{gloop}{fullcircle rotated 90 scaled (ypart vgu - ypart vgl)
                  shifted (w/2, h/2)}
  \fmfiset{vgcl}{point (1/4 * length gloop) of gloop}
  \fmfiset{vgcr}{point (3/4 * length gloop) of gloop}
  \fmfipath{prgcl,prgul,prgll,prgtl,prgbl,prgfl,prgrl}
  \fmfiset{prgcl}{vqcl -- vgcl}
  \fmfiset{prgul}{subpath ((0/4, 1/4) * length gloop) of gloop}
  \fmfiset{prgll}{subpath ((1/4, 2/4) * length gloop) of gloop}
  \fmfiset{prgtl}{vqul .. {dir 0} vgu}
  \fmfiset{prgbl}{vqll .. {dir 0} vgl}
  \fmfiset{prgfl}{vqul .. {dir 0} vgl}
  \fmfiset{prgrl}{vqll .. {dir 0} vgu}
  \fmfipath{prgcr,prgur,prglr,prgtr,prgbr,prgfr,prgrr}
  \fmfiset{prgcr}{vgcr -- vqcr}
  \fmfiset{prgur}{subpath ((3/4, 4/4) * length gloop) of gloop}
  \fmfiset{prglr}{subpath ((2/4, 3/4) * length gloop) of gloop}
  \fmfiset{prgtr}{vgu {dir 0} .. vqur}
  \fmfiset{prgbr}{vgl {dir 0} .. vqlr}
  \fmfiset{prgfr}{vgl {dir 0} .. vqur}
  \fmfiset{prgrr}{vgu {dir 0} .. vqlr}
}
  
\newcommand{\setupggH}{%
  \fmfi{dashes}{(0.5w,0.1h)--(0.5w,0.9h)}
  \fmfipath{loop}
  \fmfiset{loop}{fullcircle scaled 0.5w shifted c}
  \fmfipath{projt,projb}
  \fmfipair{Hl,Hr}
  \fmfiset{projt}{subpath ((0/2,1/2)*length loop) of loop}
  \fmfiset{projb}{subpath ((1/2,2/2)*length loop) of loop}
  \fmfiset{Hr}{point (0/2*length loop) of loop}
  \fmfiset{Hl}{point (1/2*length loop) of loop}}
\newcommand{\setupggHamp}{%
  \fmfipath{loop}
  \fmfiset{loop}{halfcircle rotated -90 scaled w shifted (0,w)}
  \fmfipath{projt,projb}
  \fmfipair{H}
  \fmfiset{projt}{subpath ((1/2,2/2)*length loop) of loop}
  \fmfiset{projb}{subpath ((0/2,1/2)*length loop) of loop}
  \fmfiset{H}{point (1/2*length loop) of loop}}
\newcommand{\setupggHH}{%
  \fmfi{dashes}{(0.5w,0.1h)--(0.5w,0.9h)}
  \fmfipath{loop}
  \fmfiset{loop}{fullcircle rotated 45 scaled 0.5w shifted c}
  \fmfipath{propl,propr,projt,projb}
  \fmfipair{Hlt,Hlb,Hrt,Hrb}
  \fmfiset{projt}{subpath ((0/4,1/4)*length loop) of loop}
  \fmfiset{propl}{subpath ((1/4,2/4)*length loop) of loop}
  \fmfiset{projb}{subpath ((2/4,3/4)*length loop) of loop}
  \fmfiset{propr}{subpath ((3/4,4/4)*length loop) of loop}
  \fmfiset{Hrt}{point (0/4*length loop) of loop}
  \fmfiset{Hlt}{point (1/4*length loop) of loop}
  \fmfiset{Hlb}{point (2/4*length loop) of loop}
  \fmfiset{Hrb}{point (3/4*length loop) of loop}}
\newcommand{\setupggHHamp}{%
  \fmfipath{loop}
  \fmfiset{loop}{halfcircle rotated -90 scaled w shifted (0w,0.5h)}
  \fmfipath{propr,projt,projb}
  \fmfipair{Hrt,Hrb}
  \fmfiset{projb}{subpath ((0/4,1/4)*length loop) of loop}
  \fmfiset{propr}{subpath ((1/4,3/4)*length loop) of loop}
  \fmfiset{projt}{subpath ((3/4,4/4)*length loop) of loop}
  \fmfiset{Hrb}{point (1/4*length loop) of loop}
  \fmfiset{Hrt}{point (3/4*length loop) of loop}}
\newcommand{\setupgggH}{%
  \fmfi{dashes}{(0.5w,0.1h)--(0.5w,0.9h)}
  \fmfipath{loop}
  \fmfiset{loop}{fullcircle scaled 0.5w shifted c}
  \fmfipath{projt,projb,projc}
  \fmfipair{Hl,Hr}
  \fmfiset{Hr}{point (0/2*length loop) of loop}
  \fmfiset{Hl}{point (1/2*length loop) of loop}
  \fmfiset{projt}{subpath ((0/2,1/2)*length loop) of loop}
  \fmfiset{projb}{subpath ((1/2,2/2)*length loop) of loop}
  \fmfiset{projc}{Hr -- Hl}}
\fmfcmd{%
  numeric joindiameter;
  joindiameter := 7thick;}
\fmfcmd{%
  vardef sideways_at (expr d, p, frac) =
    save len; len = length p;
    (point frac*len of p) shifted ((d,0) rotated (90 + angle direction frac*len of p))
  enddef;
  secondarydef p sideways d =
    for frac = 0 step 0.01 until 0.99:
      sideways_at (d, p, frac) ..
    endfor
    sideways_at (d, p, 1)
  enddef;
  secondarydef p choptail d =
   subpath (ypart (fullcircle scaled d shifted (point 0 of p) intersectiontimes p), infinity) of p
  enddef;
  secondarydef p choptip d =
   reverse ((reverse p) choptail d)
  enddef;
  secondarydef p pointtail d =
    fullcircle scaled d shifted (point 0 of p) intersectionpoint p
  enddef;
  secondarydef p pointtip d =
    (reverse p) pointtail d
  enddef;
  secondarydef pa join pb =
    pa choptip joindiameter .. pb choptail joindiameter
  enddef;
  vardef cyclejoin (expr p) =
    subpath (0.5*length p, infinity) of p join subpath (0, 0.5*length p) of p .. cycle
  enddef;}
\fmfcmd{%
  style_def double_line_arrow expr p =
    save pi, po; 
    path pi, po;
    pi = reverse (p sideways thick);
    po = p sideways -thick;
    cdraw pi;
    cdraw po;
    cfill (arrow (subpath (0, 0.9 length pi) of pi));
    cfill (arrow (subpath (0, 0.9 length po) of po));
  enddef;}
\fmfcmd{%
  style_def double_line_arrow_beg expr p =
    save pi, po, pc; 
    path pi, po, pc;
    pc = p choptail 7thick;
    pi = reverse (pc sideways thick);
    po = pc sideways -thick;
    cdraw pi .. p pointtail 5thick .. po;
    cfill (arrow pi);
    cfill (arrow po);
  enddef;}
\fmfcmd{%
  style_def double_line_arrow_end expr p =
    save pi, po, pc; 
    path pi, po, pc;
    pc = p choptip 7thick;
    pi = reverse (pc sideways thick);
    po = pc sideways -thick;
    cdraw po .. p pointtip 5thick .. pi;
    cfill (arrow pi);
    cfill (arrow po);
  enddef;}
\fmfcmd{%
  style_def double_line_arrow_both expr p =
    save pi, po, pc; 
    path pi, po, pc;
    pc = p choptip 7thick choptail 7thick;
    pi = reverse (pc sideways thick);
    po = pc sideways -thick;
    cdraw po .. p pointtip 5thick .. pi .. p pointtail 5thick .. cycle;
    cfill (arrow pi);
    cfill (arrow po);
  enddef;}
\fmfcmd{%
  style_def double_arrow_parallel expr p =
    save pi, po; 
    path pi, po;
    pi = p sideways thick;
    po = p sideways -thick;
    save li, lo;
    li = length pi;
    lo = length po;
    cdraw pi;
    cdraw po;
    cfill (arrow pi);
    cfill (arrow po);
  enddef;}
\fmfcmd{%
  style_def double_arrow_crossed_beg expr p =
    save lp;  lp = length p;
    save pi, po; 
    path pi, po;
    pi = p sideways thick;
    po = p sideways -thick;
    save li, lo;
    li = length pi;
    lo = length po;
    cdraw subpath (0, 0.1 li) of pi .. subpath (0.3 lo, lo) of po;
    cdraw subpath (0, 0.1 lo) of po .. subpath (0.3 li, li) of pi;
    cfill (arrow pi);
    cfill (arrow po);
  enddef;}
\fmfcmd{%
  style_def double_arrow_crossed_end expr p =
    save lp;  lp = length p;
    save pi, po; 
    path pi, po;
    pi = p sideways thick;
    po = p sideways -thick;
    save li, lo;
    li = length pi;
    lo = length po;
    cdraw subpath (0, 0.7 li) of pi .. subpath (0.9 lo, lo) of po;
    cdraw subpath (0, 0.7 lo) of po .. subpath (0.9 li, li) of pi;
    cfill (arrow pi);
    cfill (arrow po);
  enddef;}
\newcommand{\G}[1]{%
  \fmfi{dots,label=\footnotesize$\frac{-1}{N}$,label.dist=4thick}{#1}}
\newcommand{\A}[1]{%
  \fmfi{dbl_dots,label=\footnotesize$N$,label.dist=4thick}{#1}}
\newcommand{\C}[1]{%
  \fmfi{double_line_arrow_both,label=\footnotesize$N$,label.dist=5thick}{#1}}
\newcommand{\Gwide}[1]{%
  \fmfi{dots,label=\footnotesize$\mbox{}\qquad\frac{-1}{N}$,label.dist=4thick}{#1}}
\newcommand{\Awide}[1]{%
  \fmfi{dbl_dots,label=\footnotesize$\mbox{}\qquad N$,label.dist=4thick}{#1}}
\newcommand{\Cwide}[1]{%
  \fmfi{double_line_arrow_both,label=\footnotesize$\mbox{}\qquad N$,label.dist=5thick}{#1}}
\newcommand{\g}[2]{\fmfi{#1}{#2}}
\newcommand{\Hgg}[1]{\fmfiv{d.shape=circle,d.size=3thick,d.filled=full}{#1}}
\newcommand{\HGG}[1]{%
  \fmfiv{d.shape=circle,d.size=3thick,d.filled=full,%
         label=\footnotesize$N$,label.dist=2thick}{#1}}
\newcommand{\HGGr}[1]{%
  \fmfiv{d.shape=circle,d.size=3thick,d.filled=full,%
         label=\footnotesize$N$,label.dist=2thick}{#1}}
\newcommand{\HCA}[1]{%
  \fmfiv{d.shape=circle,d.size=3thick,d.filled=full,%
         label=\footnotesize$\frac{1}{N}$,label.dist=2thick}{#1}}

\baselineskip20pt   
\begin{flushright}
  DESY-12-103 \\ 
  SI-HEP-2012-11
\end{flushright}
\title{%
 QCD in the Color-Flow Representation
}
\author[a]{W.~Kilian,}
\author[b]{T.~Ohl,}
\author[c]{J.~Reuter}
\author[d]{and C.~Speckner}

\emailAdd{kilian@physik.uni-siegen.de}
\emailAdd{ohl@physik.uni-wuerzburg.de}
\emailAdd{juergen.reuter@desy.de}
\emailAdd{christian.speckner@physik.uni-freiburg.de}

\affiliation[a]{Universit\"at Siegen, Department Physik,
      Walter-Flex-Str.~3, 57068~Siegen, Germany}
\affiliation[b]{Universit\"at W\"urzburg, Institut f\"ur Theoretische Physik und Astrophysik,
      Emil-Hilb-Weg~22, 97074~W\"urzburg, Germany}
\affiliation[c]{DESY Theory Group, Notkestr.~85, 22603~Hamburg, Germany}
\affiliation[d]{Universit\"at Freiburg, Physikalisches Institut,
      Hermann-Herder-Str.~3, 79104~Freiburg, Germany}

\abstract{%
  For many practical purposes, it is convenient to formulate unbroken 
  non-abelian gauge theories like QCD in a color-flow basis.  We present
  a new derivation of $\mathrm{SU}(N)$ interactions in the color-flow basis by
  extending the gauge group to $\mathrm{U}(N)\times \mathrm{U}(1)'$ in such a way that the
  two $\mathrm{U}(1)$ factors cancel each other.  We use the quantum action
  principles to show the equivalence to the usual basis to all orders
  in perturbation theory.  We extend the known Feynman rules to exotic
  color representations (e.\,g.~sextets) and discuss practical
  applications as they occur 
  in automatic computation programs.}
\maketitle

\section{Introduction}

The analysis of particle physics experiments at colliders depends on
reliable theoretical predictions for cross sections of scattering
processes.  In the LHC era, hard processes with more external partons
than ever have become accessible, and their analysis is
essential for unveiling the physics at the Terascale.  This results in
a two-sided challenge: on one hand, complex Standard Model processes
must be computed with unprecedented precision and on the other hand,
the parameter space for many models for physics beyond the Standard
Model must be scanned with sufficient accuracy.  This situation
has prompted the development of computer programs that can
automatically compute differential cross sections and sample them
efficiently on phase space starting from the specification of a
Lagrangian or, equivalently, a set of Feynman rules.

In QCD, quarks and antiquarks come in three colors and
gluons in eight.  Thus, there are many amplitudes with different color
quantum numbers that must be computed and summed in quadrature,
eventually.  In Feynman diagram based calculations, the contribution
of each diagram factorizes into the dependence on color and the
dependence on flavor, polarization and momenta.  This allows to
compute the color factors once and for all.  However, for processes
with many external particles, the number of Feynman diagrams grows
factorially, and more efficient, e.\,g.~recursive, algorithms must be
used that can take advantage of the cancellations among diagrams.
Unfortunately, the color dependence does not factorize for the whole
scattering amplitude and new expansions, like color ordered amplitudes
must be employed to allow a separate computation of color factors.

It is therefore worthwhile to investigate efficient representations of
the color dependence that work for complete amplitudes, i.\,e.~sums of
Feynman diagrams.  It turns out that expressing everything in
reducible tensor products of the fundamental representation and its
conjugate instead of higher irreducible representations is beneficial
both for computation and for interfacing to parton showers,
fragmentation and hadronization.  If there are no interaction vertices
with exotic color structures, the resulting amplitudes can always be
decomposed into weighted sums of products of Kronecker-$\delta$s, so
called \emph{color flows}\footnote{If there are couplings like a totally
antisymmetric vertex
$\sum_{i,j,k=1}^3\epsilon_{ijk}\phi_i^{\vphantom{\prime}}\phi_j'\phi_k''$
in~$\mathrm{SU}(3)$, the corresponding invariant tensors can also appear
in the final result.}.

The adjoint representation of~$\mathrm{U}(N)$ is isomorphic to the
tensor product $N\times\bar{N}$ of the fundamental representation and
its conjugate. Thus, the color coordinate~$a$ of a
$\mathrm{U}(N)$-gauge boson in the adjoint representation can be
represented equivalently by a pair of~$(i,j)$ using the
decomposition~$A_{i}^{\hphantom{i}j}=A^a[T^a]_{i}^{\hphantom{i}j}$.
This equivalence is employed in 't\,Hooft's double line
notation~\cite{'tHooft:1973jz}, in which each $\mathrm{U}(N)$-gauge
boson line is replaced by a pair of matter 
and anti-matter lines for the purpose of computing color factors:
\begin{equation*}
\parbox{25\unitlength}{%
  \includegraphics{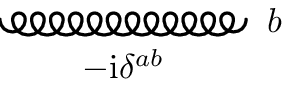}%
}\qquad\Longleftrightarrow\qquad
\parbox{25\unitlength}{%
  \includegraphics{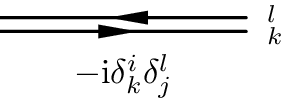}%
}\qquad.\end{equation*}
It was soon realized that the diagrammatic rules could be extended to
accomodate the tracelessness of the generators
of~$\mathrm{SU}(N)$~\cite{Canning:1975zr}.  Indeed, an efficient
diagrammatical algorithm for computing arbitrary color factors in any
gauge group was constructed~\cite{Cvitanovic:1976am} using a
generalization of the double line notation to higher
representations\footnote{%
 Independently of Feynman diagrams and quantum field theory, Penrose
 had earlier introduced a diagrammatical notation for general tensor
 calculus, including group representations~\cite{penrose}.
 This approach has evolved into the backbone of an unconventional
 textbook on the exceptional Lie algebras~\cite{Cvitanovic:2008zz}.}.

However, the color-flow representation is more than a useful
calculational device: Feynman diagrams in the double line notation
bear a striking resemblance to dual
diagrams~\cite{Harari:1981nn,Rosner:1981np} and indeed the
$N\to\infty$-limit is related to a theory of open
strings~\cite{'tHooft:1973jz}.  It is therefore no accident that the
successful semiclassical models for the fragmentation of quarks and
gluons that are based on string dynamics~\cite{Andersson:1979ij}
require a specification of the parton cross sections in the color-flow
basis~\cite{Gustafson:1982ws}.  The earliest examples of Monte-Carlo
event generators that combined the partonic cross sections for the hard
interactions with QCD-inspired fragmentation
models~\cite{Bengtsson:1982jr} provided a template for the interfacing
of hard interactions with fragmentation.   Similarly,  a color-flow
representation is used in leading order in~$1/N$ for the
implementation of color coherence in parton
showers~\cite{Marchesini/Webber}.
These interfaces became the standard
\texttt{HEPEVT} common block~\cite{HEPEVT}, which includes color flow
information. Originally developed for interfacing Monte-Carlo event
generators for hard scattering matrix elements with libraries or
external programs for parton showers, fragmentation and hadronization
at LEP1, it was widely adopted and has grown into the Les
Houches Accords~(LHA)~\cite{Boos:2001cv}.

Therefore the color-flow basis is particularly suited for the
computation of partonic scattering amplitudes that will be interfaced
with the other components of a Monte-Carlo event generator.  Indeed,
most generators for hard multi-parton scattering can provide the
corresponding information.  One can, however, go one step further and
not only represent the final result in a color-flow basis, but use
Feynman rules in a color-flow basis already for the computation.  A
lucent derivation of these Feynman rules was given
in~\cite{Maltoni:2002mq}.  Since the derivation
in~\cite{Maltoni:2002mq} is based on the QCD Lagragian, it is not
necessary to use Feynman rules at all, but one can implement modern
recursive
algorithms~\cite{Caravaglios:1995cd,Kanaki:2000ey,Moretti:2001zz}
directly in the color-flow
basis~\cite{Kilian:2007gr,Cafarella:2007pc,Hagiwara:2010vk}.

However, the derivation~\cite{Maltoni:2002mq} of the color-flow
representation is incomplete in two directions: firstly, there is no
consideration of interactions with more exotic color structures, in
particular beyond QCD with fermionic matter in the fundamental
representation, and secondly the discussion is deliberately
confined to tree level amplitudes.  These two limitations are in fact
related, and overcoming them is not only of theoretical interest: the most
important light Higgs production channel at LHC involves the
dimension-5 operator~$H\,\tr(F_{\mu\nu}F^{\mu\nu})$, which arises from
a loop and corresponds to an octet-octet-singlet coupling that cannot
be described straightforwardly in the framework provided
by~\cite{Maltoni:2002mq}.
Indeed, our generalization of the color-flow representation was
prompted by the implementation of the effective Higgs-gluon-gluon
vertex in WHIZARD's~\cite{Kilian:2007gr} optimized matrix element
generator~O'Mega~\cite{Moretti:2001zz}.  While it was simple to
``fudge'' the Feynman rules for a single insertion, it turned out that
these prescriptions would give incorrect results for multiple
insertions.

This paper is organized as follows: we start with a motivation and
description of our implementation of color-flow QCD in
section~\ref{sec:model} and discuss quantum field-theoretical aspects
in more detail in section~\ref{sec:qft}.  In
section~\ref{sec:renormalization}, we show that our description is
valid to all orders in perturbation theory.  In the following two
sections we discuss some applications in detail: tree level amplitudes
in section~\ref{sec:trees} and the already mentioned effective
interactions from loops in section~\ref{sec:loops}.  For reference, we
repeat the color-flow Feynman rules for QCD with fermionic matter in
the fundamental representation in appendix~\ref{sec:rules}.  As an
example of exotic color representations, we extend
those rules to color-sextet particles in appendix~\ref{sec:exotics}.

\section{The Model}
\label{sec:model}

In this section, we construct a variant of QCD which contains two
extra gauge degrees of freedom.  The model is set up not for its
physics content, but for its usefulness in practical perturbative
calculations.  In fact, as we will show in later sections, in physical
quantities the extra degrees of freedom cancel, so all predictions are
identical to ordinary QCD.

\subsection{QCD Preliminaries}

We consider QCD with a single fermionic matter species in the
$N$-dimensional fundamental representation of $\mathrm{SU}(N)$.  While we are
really interested in the $N=3$ case, it is sometimes useful to
explicitly keep the dependence on $N$, as we will do through most of
this paper.  Furthermore, we will not assume that the matter representation
is vector-like as in QCD, although we do follow the QCD notation with
Dirac fermions in the Lagrangian.

Let us first recall basic facts of perturbative QCD.  The perturbation
series is derived from a Lagrangian which splits into two parts,
\begin{equation}
  \LL = \LL_{\text{inv}} + \LL_{\text{gf}},
\end{equation}
a gauge-invariant part $\LL_{\text{inv}}$ and a gauge-fixing part
$\LL_{\text{gf}}$.  The gauge-invariant part is given by\footnote{We
  set the masses of matter fields to zero, for brevity.}
\begin{equation}
\label{QCD}
  \LL_{\text{inv}} =
    -\frac{1}{2g^2}\tr\, \hat G_{\mu\nu} \hat G^{\mu\nu}
    +\bar\psi\left(\ii\fmslash\pd + \hat {\fmslash A}\right)\psi.
\end{equation}
The gauge-fixing part depends on the chosen gauge-fixing procedure.
In a manifestly covariant formulation with linear gauge-fixing, it
takes the form
\begin{equation}
\label{QCD-gf}
  \LL_{\text{gf}} = \frac{2}{g^2}\tr \hat B(\pd\cdot \hat A)
                  + \frac{\xi}{g^2}\tr \hat B^2
                  - \frac{2}{g^2}\tr\hat{\bar c}\pd^\mu(\pd_\mu \hat c 
                                                         - \ii[\hat A_\mu,\hat c]).
\end{equation}
Since the matter fields are in the fundamental (defining)
$N$-dimensional representation of $\mathrm{SU}(N)$, we choose to represent the
Lie-algebra valued fields $G,A,B,c,\bar c$ by traceless $N\times N$
matrices.\footnote{For later convenience, we mark traceless matrix
  fields by a hat, $\hat A_\nu$.}
The field-strength tensor $\hat G$ can be expressed in terms of the
gauge potential $\hat A$ as
\begin{equation}
  \hat G_{\mu\nu} 
  = \pd_\mu \hat A_\nu - \pd_\nu \hat A_\mu - \ii[\hat A_\mu,\hat A_\nu].
\end{equation}
Furthermore, the gauge-fixing term involves
a Fadeev-Popov~\cite{Faddeev:1967fc} ghost field $\hat c$,
an antighost field $\hat {\bar c}$ and
a Nakanishi-Lautrup (NL)~\cite{Nakanishi/Lautrup}
auxiliary field $\hat B$. The latter can be integrated
out in order to obtain the more familiar form of the gauge fixing Lagrangian
\begin{equation}
  \LL_{\text{gf}} = - \frac{1}{g^2\xi}\tr (\pd\cdot \hat A)^2
                  - \frac{2}{g^2}\tr\hat{\bar c}\pd^\mu(\pd_\mu \hat c 
                                                         - \ii[\hat A_\mu,\hat c]).
\end{equation}

We may assume a manifestly gauge-invariant renormalization procedure such as
$\overline{\mathrm{MS}}$, so that the Lagrangian retains the
form~(\ref{QCD},\ref{QCD-gf}) in each order of the perturbative
expansion.\footnote{A generic renormalization procedure could make 
  non-invariant terms in the renormalized Lagrangian necessary that ensure gauge
  invariance of the effective action.}  The fields in the Lagrangian are
understood to be renormalized, order by order, such that the model contains
only two (renormalized) real parameters, the gauge-coupling $g$ and the
gauge-fixing parameter $\xi$.  The latter drops out of physical quantities.

The gauge group $\mathrm{SU}(N)$ is a subgroup of the general linear group
$\mathrm{GL}(N)$, therefore the fields (e.\,g., $\hat A$), in the $N\times
N$ matrix representation, obey algebraic
constraints:
\begin{equation}
\label{Aconstraints}
  \hat A^\dagger = \hat A
\qquad\text{and}\qquad
  \tr\hat A = 0.  
\end{equation}
The constraints are automatically satisfied if we introduce the usual
basis for the Lie algebra representation, $T^a$ ($a=1,\ldots N^2-1$),
and write $\hat A = \sum_a A^aT^a$.  The basis elements obey the
hermiticity and trace conditions:
\begin{equation}
  \tr\, T^a = 0,
  \qquad
  \tr\, T^aT^b = \tfrac12\delta^{ab},
  \qquad
  [T^a, T^b] = if^{abc}T^c,
  \qquad
  (T^a)^\dagger = T^a
\end{equation}
with the real structure constants $f^{abc}$.  Each Feynman graph can
be factorized into a kinematical and a color amplitude.  The color
amplitude consists of a string of $T$ and $f$ tensors, contracted
over all internal indices, and representable as a combination of
tensors with open external indices $a,b,\ldots,i,j,\ldots$.  For squared
amplitudes, all indices are summed over, hence the particular
representation becomes irrelevant.  Various schemes exist to compute
color amplitudes exactly and efficiently.

\subsection{The Color-Flow Representation}

In the following, we will eliminate the basis $T^a$, and instead work
with the individual matrix elements of the gauge fields in the chosen
Lie algebra representation~\cite{Cvitanovic:2008zz}.  This approach is
known as a \emph{color-flow representation}.  For concreteness, we
consider the gauge potential $\hat A$.  Without algebraic constraints,
a $N\times N$ matrix field contains $N^2$ complex degrees of freedom.

We have to implement the algebraic contraints~(\ref{Aconstraints}).
The hermiticity condition is
\begin{equation}\label{eq:Aij}
  \hat A^i{}_j = (\hat A^\ast)^j{}_i.
\end{equation}
In effect, each unordered index combination $(ij)$ carries only one,
instead of two, complex degree of freedom.  We can understand this as
if each \emph{ordered} index combination $(i,j)$ carried one
\emph{real} degree of freedom.  For the propagator in a Feynman graph,
a value of an index can be represented by a colored line.  There are
$N$ different colors which correspond to the $N$ values each index may
take.  An arrow of the line indicates whether the index is the lower
or upper one.  Hence, $\hat A$ propagators carry two color lines with
opposing directions.

Similarly, matter propagators are represented by a single directed
color line.  External states are represented by a terminating line
(matter) or two opposing terminating lines (gauge).  As long as we
consider only the interactions following from the Lagrangian as quoted
above, the rule of matrix multiplication trivially ensures that color
is conserved at each vertex, i.\,e., lines end only at external states.

While this graphical approach is straightforward, we also have to
enforce the trace condition from~(\ref{Aconstraints}), explicitly
\begin{equation}\label{eq:Aii}
  \sum \hat A^i{}_i = 0,
\end{equation}
which reduces the number of degrees of freedom by one.  The
implementation of this constraint amounts to additional modified
color-flow patterns in Feynman graphs~\cite{Maltoni:2002mq}.  While
the construction of these patterns is easy to understand, they are, by
definition, a non-local modification of the naive color-flow
expansion.  This is a complication in automatic, in particular
non-diagrammatic, algorithms, which we want to avoid.

Instead, we follow a field-theoretical approach and set up a modified
QCD theory.  This theory has the algebraic constraints inherent in its
field content, such that they are automatically satisfied by the
generated color-flow amplitudes.  In essence, it generates (i) terms
that complete the $A$ field to a $\mathrm{U}(N)$ matrix so that it has no trace
condition and naive color-flow Feynman rules apply, and separately
(ii) terms that subtract the spurious contributions without
complicating the color-flow rules.  To make the diagrammatic expansion
unambiguous, the two fields have to be formally independent of each
other.

\subsection{Singlet and Phantom Gluons}

We denote the field which complements the gluon field matrix as the
\emph{singlet gluon}~$A_0$.  This field is associated with a
``zeroth'' generator $T^0$ which is not traceless but satisfies
\begin{equation}
  \label{eq:trace_norm}
  \tr\, T^0 = \sqrt{\tfrac{N}{2}},
  \qquad
  \tr\, T^aT^0 = \tfrac12\delta^{a0},
  \qquad
  [T^a,T^0] = 0,
\end{equation}
and let $a$ run from $0$ to $N^2-1$.  Explicitly,
\begin{equation}
  (T^0)^i_{\;j}=\frac{1}{\sqrt{2N}}\,\delta^i_j,
\end{equation}
(We choose upper indices for the fundamental and lower indices for the
anti-fundamental color representation.)  The matrix field resulting
from combining $\hat A$ with $A_0$
\begin{equation}
  A = \hat A + A^0 T^0 = \left(\sum_{a=0}^{N^2-1} A^a T^a\right)
\end{equation}
has $N^2$ independent components, the color-flow gluons $A^i_{\hphantom{i}j}$.
Analogously, for the NL and ghost fields, we replace the traceless
fields by unconstrained fields $B,c,\bar c$ and singlet fields
$B_S,c_S,\bar c_S$, respectively.  The field strength $G$ is modified
accordingly.

For the subtraction terms, we introduce a \emph{phantom gluon} field
$\tilde A$.  The phantom gluon is an independent $\mathrm{U}(1)$ gauge boson
with the wrong sign in the propagator.  Like the other gluons, it
couples to the matter fields via the QCD coupling $g$.  It does not
couple to gluon fields.  We also introduce the corresponding NL field
$\tilde B$.  (We may also introduce phantom ghost fields, but these
can be omitted with linear gauge fixing, since the $\mathrm{U}(1)$ gauge group
is abelian.)

\subsection{Color-Flow QCD}

We can now set up the Lagrangian for the color-flow version of QCD,
that includes the singlet gluon (as a component of the gluon matrix
$A^i_{\hphantom{i}j}$) and the phantom gluon $\tilde A$, together with their
associated field strength, NL and ghost fields.

The Lagrangian\footnote{For the purpose of discussing renormalization,
  we incorporate the gauge coupling in the normalization of the
  gauge-field and gauge-fixing
  terms.  To derive Feynman rules later, we will apply a trivial field
  renormalization to transfer $g/\sqrt{2}$ to the vertices and obtain
  a canonically normalized $\mathrm{U}(N)$ gauge field from~\eqref{QCD-phantom}.
  However, we will retain the factor~$N$ in
  the phantom kinetic term and it will appear in the phantom propagator
  and external states.}
is
\begin{align}
\label{QCD-phantom}
  \LL  &= 
    -\frac{1}{2g^2}\tr\, G_{\mu\nu} G^{\mu\nu}
    +\frac{N}{2g^2}{\tilde G}_{\mu\nu} {\tilde G}^{\mu\nu}
    +\bar\psi\left(\ii\fmslash\pd 
                   + {\fmslash A} - {\fmslash{\tilde A}}\right)\psi
\nonumber\\
  &\quad
    + \frac{2}{g^2} \tr B(\pd\cdot A)
    - \frac{2N}{g^2} \tilde B(\pd\cdot \tilde A)
    + \frac{\xi}{g^2}\tr B^2
    - \frac{\xi N}{g^2}{\tilde B}^2
    + \LL_{\text{ghost}}
\end{align}
where
\begin{align}
\label{QCD-phantom-ghost}
  \LL_{\text{ghost}} &=
    - \frac{2}{g^2} \tr {\bar c}\pd^\mu(\pd_\mu c - \ii[A_\mu,c])
\end{align}
and the decoupling ghosts for the phantom gluon have been omitted.

A priori, this theory is \emph{different} from QCD.  We will denote
it as ``color-flow QCD'' and we will show its equivalence to QCD in
section~\ref{sec:qft}.  It contains $N^2$ instead of $N^2-1$
independent degrees of freedom in the $A$ field, plus the additional
$\mathrm{U}(1)'$ gauge field $\tilde A$ with wrong-sign propagator.  Thus, it
can be regarded as a $\mathrm{U}(N)\times \mathrm{U}(1)'$ or, equivalently, $\mathrm{SU}(N)\times
\mathrm{U}(1)\times \mathrm{U}(1)'$ gauge theory.  In a path-integral formulation, there
are independent integrations over all $N^2+1$ gauge components.  The
singlet field $A^0$ is present, but hidden in the gauge-field matrix.

In the Lagrangian~(\ref{QCD-phantom}), we have encoded the algebraic
properties of the color-flow decomposition as a (perturbative) quantum
field theory on its own.  Therefore, we can expect that amplitudes
calculated from this field theory manifestly exhibit the advantages of
this approach.

In calculations, we therefore retain the decomposition of the
matrix-valued fields in terms of their matrix elements, $A^i_{\hphantom{i}j}$.  As
explained above, these non-hermitean vector bosons are represented by
double lines, one for the color and one for the anticolor flow,
flowing in opposite directions.  The vertex color factors are
Kronecker deltas, which diagrammatically correspond to color lines
continuing through the vertex.  The phantom gluon, on the other hand,
is analogous to a photon.  In particular, it carries no color flow.
Color lines always terminate at external states.  Hence, for any tree
diagram, the color factor is simply $N^p$, where $p$ is the number of
distinct color lines in the diagram (after squaring).  Since this is
determined by the colors of the external states, the whole color
algebra becomes trivial.  The price for this is a proliferation of
extra Feynman diagrams (which have identical kinematics, however, and
thus do not blow up the calculation.)

We list the Feynman rules that follow from~(\ref{QCD-phantom}) in
Appendix~\ref{sec:rules}.

\subsection{Example}

Let us look at the exchange of a gluon between fermion lines.
In ordinary QCD, this is represented by the diagram
\begin{align*}
\parbox{30\unitlength}{%
  \includegraphics{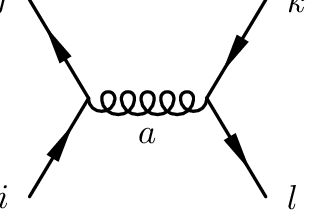}%
}
\end{align*}
\vspace{\baselineskip}
In color-flow QCD, we get two diagrams, one for gluon and one for
phantom exchange
\begin{align*}
\parbox{30\unitlength}{%
  \includegraphics{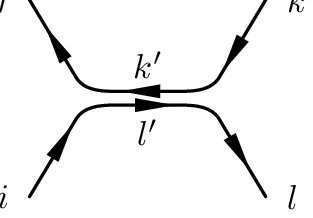}%
\\[0.5\baselineskip]
\centerline{$\frac{1}{\sqrt2}\delta_i^{l'}\delta_{k'}^j\;\;
  \frac{1}{\sqrt{2}} \delta^l_{l'} \delta^{k'}_k$}
}
&&
\parbox{30\unitlength}{%
  \includegraphics{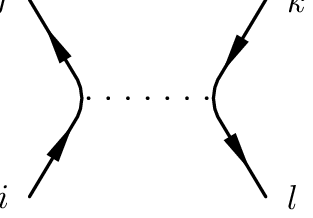}%
\\[0.5\baselineskip]
\centerline{$\frac{-1}{\sqrt{2}}\delta_i^j\;\;\left(\frac{-1}{N}\right)\;\;\frac{-1}{\sqrt{2}}\delta_k^l$}
}
\end{align*}
\vspace{\baselineskip}
with the correct sum
\begin{equation}
\label{TaTa}
  \frac12\left(\delta_i^l\delta_k^j -
    \frac{1}{N}\delta_i^j\delta_k^l\right)
  = \sum_a (T^a)_i^j (T^a)_k^l.
\end{equation}
Usually, this color-flow decomposition is obtained by applying the
relation~(\ref{TaTa}) backwards.  The Feynman rules derived from the
color-flow Lagrangian generate it directly.

Also at the squared-amplitude level, the color-flow representation
leads to a reorganization of the color-flow calculation, which
considerably simplifies the algorithm.  There is a further benefit:
common parton-shower and hadronization algorithms expect the color
connections to be expressed in a color-flow basis.  We obtain both the
exact amplitude (at the given order) and, by omitting the interference
terms and all terms containing external phantoms, the projection on
the possible color-flow patterns in a single step.

In summary, we work with a trivial version of color algebra since the
complete color flow information is represented on a graphical basis.
Subtractions in the color factors are accounted for by extra diagrams
which involve the phantom gluon.  No approximation is involved.

\section{Field-theoretical Considerations}
\label{sec:qft}

In the following, we demonstrate that color-flow QCD as
described above is equivalent to the original $\mathrm{SU}(N)$ theory, i.\,e.,
$A^0$ and $\tilde A$ cancel each other in physical quantities.  This
equivalence may appear trivial, at least in tree-level amplitudes.
However, it bears some subtleties, in particular if it is to be
extended to all orders.  To promote the approach from a technical
device to a consistent field theory, we have to prove that the
equivalence can be maintained, by suitable renormalization conditions,
to all orders in perturbation theory.

This situation is similar to the covariant quantization of QED, where
the scalar and longitudinal photon components, which are both
unphysical, cancel in physical quantities.  In a nonabelian theory,
this requires the additional introduction of ghost
fields~\cite{Faddeev:1967fc,BRST,Kugo:1977zq,Piguet}.  In the Fock space, the
unphysical fields lead to indefinite-norm states, but it is possible
to build a Hilbert space out of equivalence classes of physical states
in which the $S$~matrix is unitary.
As we will show below, the same argument applies in the
present case: the phantom gluon generates negative-norm states, but it
is always produced coherently with the singlet gluon (which is now
hidden in the $N\times N$ gluon matrix), and therefore a unitary
$S$~matrix is well defined.

\subsection{Algebraic Identities at Lowest Order}

A consistent formulation of a quantum field theory is based on the
symmetries that the effective action, including the integrated
Lagrangian as its tree approximation, has to satisfy order by order in
a perturbative expansion.  These are representable as algebraic
identities~\cite{Piguet}.

In the present subsection, we look only at the lowest order, so we
temporarily ignore the complications that necessitate the replacement
of gauge symmetry by BRST symmetry, ghost fields, etc.  Given the
fields $A^i_{\;j}$, $\tilde A$, $B^i_{\;j}$, $\tilde B$, $\bar\psi$,
$\psi$, we observe that the lowest-order action $\Gamma^0 = \int\LL$
from~(\ref{QCD-phantom}), with the ghost term omitted,
satisfies the Ward identity\footnote{All functional derivatives are
defined to act from the left which introduces a sign for the variation
of fermions
\begin{equation*}
  \delta\Gamma
     = \delta\bar\psi_\alpha \frac{\delta\Gamma}{\delta\bar\psi_\alpha}
     + \delta\psi_\alpha \frac{\delta\Gamma}{\delta\psi_\alpha}
     = \delta\bar\psi \frac{\delta\Gamma}{\delta\bar\psi}
     - \frac{\delta\Gamma}{\delta\psi} \delta\psi\,.
\end{equation*}}
of $\mathrm{U}(N)$,
\begin{align}
\label{Ward}
  \tr\left[(\delta A)\frac{\delta\Gamma}{\delta A}\right]
  + (\delta\bar\psi)\frac{\delta\Gamma}{\delta\bar\psi}
  - \frac{\delta\Gamma}{\delta\psi}(\delta\psi) &=
   \frac{2}{g^2}\tr\left[B\pd\cdot\delta A\right],
\end{align}
where the gauge transformations of the fields are
\begin{equation}
  \delta A = \pd\omega - \ii[A, \omega]\,,
  \qquad
  \delta\tilde A = 0\,,
  \qquad
  \delta\psi = \ii\omega\psi\,,
  \qquad
  \delta\bar\psi = -\ii\bar\psi\omega\,,
  \qquad
  \delta B = 0\,.
\end{equation}
This Ward identity decomposes into separate identities for
the $\mathrm{SU}(N)$ and $\mathrm{U}(1)$ factors.
Even though we are interested only in the $\mathrm{U}(1)$ part, we
want to keep our notation free of the clutter of indices. Therefore, we
retain the matrix notation and use traces to project on
the~$\mathrm{U}(1)$.  Indeed, we can evaluate the Ward
identity~(\ref{Ward}) for $\omega = \omega_0\um$ to find
\begin{align}
\label{singlet-Ward}
  \pd_\mu\tr\frac{\delta\Gamma}{\delta A_\mu}
  + \ii\left(\bar\psi\frac{\delta\Gamma}{\delta\bar\psi}
          + \frac{\delta\Gamma}{\delta\psi}\psi\right)
  &= - \frac{2}{g^2} \pd^2\tr B\,.
\end{align}
This is accompanied by the gauge-fixing condition for the singlet
\begin{align}
\label{gauge-fixing}
  \tr\frac{\delta\Gamma}{\delta B} &= 
  \frac{2}{g^2}\left(\tr\pd\cdot A + \xi\tr B\right).
\end{align}
Furthermore we define, for any
matrix-valued field $M\in\{A, G, \ldots\}$, the linear combinations
\begin{subequations}
\label{eq:redef}
\begin{equation}
  M_\pm = \frac{1}{\sqrt2}\left(\frac{1}{{N}}\tr M \pm \tilde M\right)\\
\end{equation}
and the dual functional operators
\begin{equation}
  \frac{\delta\Gamma}{\delta M_\pm} 
    = \frac{1}{\sqrt2}\left(\tr\frac{\delta\Gamma}{\delta M}
        \pm \frac{\delta\Gamma}{\delta\tilde M}\right)
\end{equation}
\end{subequations}
with
\begin{equation}
  \frac{\delta M_\pm(x)}{\delta M_\pm(y)} = \delta^4(x-y)\,,\qquad
  \frac{\delta M_\pm(x)}{\delta M_\mp(y)} = 0\,.
\end{equation}
Using~\eqref{eq:redef}, we can rewrite terms appearing in the
Lagrangian as
\begin{subequations}
\label{eq:M+M-}
\begin{align}
   \frac{1}{2} \tr M^2 - \frac{N}{2} \tilde M^2
     &= \frac{1}{2} \tr\hat M^2 + N M_+ M_- \\
  T^0 M^0 - \tilde M &= \sqrt{2}\,M_-\,.
\end{align}
\end{subequations}

Computing the derivative of the lowest-order action with respect to $A$ and $\tilde A$ and
combining the results, we
obtain the \emph{phantom equation}
\begin{align}
\label{phantom-eq}
  \frac{\delta\Gamma}{\delta A_+{}_\nu}
  &= \frac{2N}{g^2}\left(\pd_\mu G^{\mu\nu}_- - N\pd^\nu B_-\right).
\end{align}
It expresses the cancellation between singlet and phantom terms.
Since $A_+$ is an abelian gauge field, applying $\delta/\delta
A_+^\nu$ to the ghost term~(\ref{QCD-phantom-ghost}) yields zero, so
the phantom equation holds for the complete lowest-order
Lagrangian in the BRST formalism~\cite{BRST}.

\subsection{Fock-Space Cancellation}
\label{sec:cancellations}

To clarify the physical implications of the phantom equation~\eqref{phantom-eq}, let us
use~\eqref{eq:M+M-} to express the singlet and phantom fields $(\frac{1}{N}\tr A, \tilde A)$
by $(A_+,A_-)$.  The remainder of the matrix $A$ is traceless,
this is the original $\mathrm{SU}(N)$ gluon with all of its interactions.  We
must show that the extra fields $A_+$ and $A_-$ do not contribute to
observable quantities.

We eliminate the $B$ fields via their equations of motion.  The terms in
the resulting Lagrangian that depend on $A_+$ or $A_-$ take the form
\begin{align}
\label{QCD-phantom'}
  \LL  &= 
    -\frac{N}{g^2}{G}_+{}_{\mu\nu} {G}_-{}^{\mu\nu}
    - \frac{2N}{\xi g^2} (\pd\cdot A_+)(\pd\cdot A_-)
    + \sqrt{2}\,\bar\psi\left({\fmslash A_-}\right)\psi\,.
\end{align}
The propagator interchanges $A_+$ and $A_-$, but only $A_-$ couples to
matter.  Couplings to ordinary gluons are absent in~\eqref{QCD-phantom'}, because
$G_\pm^{\mu\nu}=\partial^\mu A_\pm^\nu - \partial^\nu A_\pm^\mu$.  As a result, 
whenever a singlet/phantom gluon is created by a matter current, it
cannot be annihilated, and vice versa.  In short, $A_+$ and $A_-$ do
not introduce any observable interactions.

In fact, we may freely add any term to the effective action that
depends on $A_-$ at least linearly, but has no dependence on $A_+$.
Any such interaction is unobservable.

This pattern will continue to hold to all orders, unless a loop
diagram induces an interaction of some current with $A_+$.  However,
the phantom equation (\ref{phantom-eq}) prohibits this: it tells that
the interactions of $A_+$ are completely accounted for by the
leading-order Lagrangian, up to a renormalization of the coupling
constant.  Therefore, if we can maintain the phantom equation on the
effective action to all orders, observables computed from the
color-flow QCD theory are identical to those from ordinary QCD.

In actual applications, we do not express the Lagrangian in terms of
$A_+$ and $A_-$.  Instead, we work with the $\mathrm{U}(N)$ field $A$ and 
the $\mathrm{U}(1)'$ phantom $\tilde A$, as we did in the previous sections.
Hence, the Fock-space cancellation is not entirely trivial since it
involves cancellations among graphs that belong to different gauge
groups, at least superficially.  We have to enforce the phantom
equation explicitly in order to guarantee that no $A_+$ interaction
arises in the effective action.


\subsection{Symmetries}
\label{sec:symmetries}

If the phantom equation~(\ref{phantom-eq}) is combined with the Ward
identity~(\ref{singlet-Ward}) and the gauge-fixing
condition~(\ref{gauge-fixing}), we obtain an analogous Ward identity
and gauge-fixing condition for the phantom gluon.  The complete
symmetry, at lowest order, is $\mathrm{U}(N)\times \mathrm{U}(1)'$, with the additional
condition that all gauge couplings are equal (to the $\mathrm{SU}(N)$ gauge
coupling).

In the previous subsection, we introduced linear combinations of the
extra gauge fields, $A_+$ and $A_-$.  Correspondingly, we may identify
gauge groups $\mathrm{U}(1)_+$ and $\mathrm{U}(1)_-$ which are orthogonal combinations
of the original extra $\mathrm{U}(1)$ and $\mathrm{U}(1)'$ gauge groups.  These gauge
invariances are described by the Ward identities
\begin{align}
\label{WI-A-}
  \pd^\nu\frac{\delta\Gamma}{\delta A_-^\nu}
  + \ii\sqrt2\left(\bar\psi\frac{\delta\Gamma}{\delta\bar\psi}
                 + \frac{\delta\Gamma}{\delta\psi}\psi\right)
  &= -\frac{2N}{g^2} \pd^2B_+,
\\
\label{WI-A+}
  \pd^\nu\frac{\delta\Gamma}{\delta A_+^\nu}
  &= -\frac{2N}{g^2} \pd^2B_-,
\end{align}
which follow from the Ward identity~(\ref{singlet-Ward}) and the phantom
equation~(\ref{phantom-eq}).

The corresponding gauge-fixing conditions are
\begin{align}
\label{gauge-fixing-pm}
  \frac{\delta\Gamma}{\delta B_\pm} &= 
  \frac{2N}{g^2} \left(\pd\cdot A_\mp + \xi N B_\mp\right).
\end{align}
Note that the matter fields are neutral under $\mathrm{U}(1)_+$.  This is in
line with the observation that $A_+$ does not interact at all.

We will demonstrate in section~\ref{sec:renormalization} that
interactions of $A_+$ are
forbidden in the effective action to all higher orders, 
by the phantom equation, and there is a $\mathrm{U}(1)_+$ symmetry.
On the other hand, regarding $\mathrm{U}(1)_-$, there is no guarantee that this
symmetry can be enforced to all orders.  If the gauge representation
of the matter fields is chiral (QCD happens to be vector-like), the
extra $\mathrm{U}(1)_-$ symmetry might be anomalous even if there is no anomaly
of the original $\mathrm{SU}(N)$.  However, by the argument in the previous
subsection, we see that a $\mathrm{U}(1)_-$ anomaly will not contribute to
observables, so it does not invalidate unitarity of the $S$ matrix
projected onto physical states.


\section{Renormalization}
\label{sec:renormalization}

In this section we outline a proof that the color-flow version of QCD
can be extended consistently to all orders in perturbation theory.  We
follow the algebraic renormalization procedure, as explained in detail
in Ref.~\cite{Piguet}.  Algebraic renormalization is mainly useful for
proving the perturbative renormalizability to all orders of a class of
models, in particularly for demonstrating the absence of anomalies,
i.\,e.~that symmetries can be maintained in the quantized theory.

It is convenient in practical loop calculations to employ a
regularization that maintains the rigid and gauged symmetries of a
Lagrangian in order to reduce the number of required counterterms.  For
recursive proofs to all orders, on the other hand, it is
important to prove the absence of divergencies that would require
the introduction of counterterms which would break a symmetry of the
$n$-loop effective action.
For this purpose one can use a non-invariant regulator for
which the model-independent quantum action
principles~\cite{Lowenstein:1971jk,Lam:1972mb,Clark:1976ym,Breitenlohner:1977hr}
have been established, which limit the possible counterterms to a finite
set that can be enumerated explicitly.

The inductive proof of renormalizability proceeds then in three
stages: first one derives a set of Slavnov-Taylor~(ST) identities for
the effective action from the rigid and gauged symmetries of the model
under consideration (cf.~sect.~\ref{sec:ST1}).  Subsequently, one
proves that the tree-level effective action (i.e., the Lagrangian
integrated over space-time) is the unique solution of these functional
equations (cf.~sect.~\ref{sec:ST2}), up to field and coupling constant
renormalizations.  This establishes an induction hypothesis.  For the
induction step at $n$-loop order, one shows recursively that these
ST~identities can be maintained in order~$n+1$ of perturbation theory
through a suitable choice of local counterterms if they are satisfied
by the $n$-loop effective action (cf.~sect.~\ref{sec:AR}).

The induction step consists of several parts. One first restricts the
terms that can possibly violate the ST identities at order~$n+1$,
assuming that they hold at order~$n$.  The quantum action principles
guarantee that these terms take the form of local operators with
well-defined dimension and quantum numbers, and the ST identities
impose further algebraic constraints.  The coefficients of the local
operators can be computed as the values of certain Feynman graphs.  If
there are operators with nonvanishing coefficients, one has to check
whether they can be cancelled by adding suitable non-invariant
counterterms to the Lagrangian.  This is a purely algebraic problem.
If there is a solution, the ST identities can be restored by the
counterterms and renormalizability is proved.  Otherwise, the symmetry
is manifestly broken by an anomaly.

\subsection{Conditions Imposed on the Effective Action}
\label{sec:ST1}

First, we need a precise (perturbative) definition of the quantum
field theory.  We define the effective action of color-flow QCD as a
solution of the usual ST identity for $\mathrm{SU}(N)$, here
extended to $\mathrm{U}(N)$:
\begin{align}
\label{ST}
  \int d^4x\left(\tr\left[\frac{\delta\Gamma}{\delta\rho_A}\,
                 \frac{\delta\Gamma}{\delta A}\right]
               + \frac{\delta\Gamma}{\delta\rho_\psi}\,
                 \frac{\delta\Gamma}{\delta\bar\psi}
               + \frac{\delta\Gamma}{\delta\psi}\,
                 \frac{\delta\Gamma}{\delta\bar\rho_\psi}
               + \tr\left[\frac{\delta\Gamma}{\delta\rho_c}\,
                 \frac{\delta\Gamma}{\delta c}\right]
               + \tr\left[B\frac{\delta\Gamma}{\delta\bar c}\right]
            \right) = 0
\end{align}
with the linear gauge-fixing equation
\begin{align}
\label{gf}
  \frac{\delta\Gamma}{\delta B}
     = \frac{2}{g^2}\left(\pd\cdot A + \xi B\right).
\end{align}
These identities involve the gauge, gauge-fixing, and ghost fields
$A^i_j,B^i_j,c^i_j,\bar c^i_j$, expressed in the color-flow basis, the
matter fields $\psi,\bar\psi$, and the sources for the BRST variations,
$\rho_A,\rho_\psi,\bar\rho_\psi,\rho_c$.  

Furthermore, the effective action depends on the phantom field $\tilde
A$ and its associated NL field $\tilde B$.  The dependence on those
fields is accounted for by the phantom equation~(\ref{phantom-eq}),
which we state here in the form
\begin{equation}
\label{phantom-eq-exp}
  \tr\frac{\delta\Gamma}{\delta A^\nu}
  + \frac{\delta\Gamma}{\delta\tilde A^\nu}
  - \frac{2}{g^2}\left(\pd^\mu \tr G_{\mu\nu} -
     N \pd^\mu\tilde G_{\mu\nu}\right)
  + \frac{2}{g^2}\left(\pd_\nu\tr B -  N \pd_\nu\tilde B\right) 
  = 0\,,
\end{equation}
and by the gauge-fixing condition for $\tilde A$,
\begin{align}
\label{gf-phantom}
  \frac{\delta\Gamma}{\delta\tilde B} 
  = - \frac{2N}{g^2}\left(\pd\cdot\tilde A + \xi\tilde B\right).
\end{align}

The algebraic conditions are supplemented by renormalization
conditions which fix the field normalizations and a minimal set of
interaction parameters.  In the present model, these are just the
gauge couplings.

Alternatively, in a minimal-subtraction scheme such as
$\overline{\textrm{MS}}$, we read off renormalization conditions after
regularizing the loop integrals and subtracting the divergencies.
This can also be done here, but with a caveat: we
may use any renormalization scheme for the $\mathrm{SU}(N)$ coupling and for
the ordinary $\mathrm{SU}(N)$ fields, but in order to keep the simplicity of
the approach, we impose the condition that the singlet $\mathrm{U}(1)$
coupling is equal to the $\mathrm{SU}(N)$ coupling and that the field
normalizations are identical.  (The ST identity does not enforce
this 
condition since $\mathrm{U}(N)$ is not simple.) For the $\mathrm{U}(1)'$ coupling and the
phantom field, everything is then fixed by the phantom equation and
the phantom gauge-fixing condition.

\subsection{Lowest-Order Solution}
\label{sec:ST2}

The next step in the renormalization procedure is to identify the
lowest-order effective action.  We can verify that the
Lagrangian~(\ref{QCD-phantom}) is the unique solution of the
constraints of the preceding section (with correct gauge group and
representation) which is an integrated local polynomial in the fields
of dimension equal to four.

The BRST invariance of the theory, and of the lowest-order
approximation in particular, is encoded in the ST identity~(\ref{ST})
and leads to lowest-order BRST transformations of the form
\begin{align}
  sA &= \pd c - \ii[A,c]
&
  sc &= \ii c c
&
  s\bar c &= B
&
  sB &= 0
\nonumber\\
&&
  s\psi &= ic\psi
&
  s\bar\psi &= \ii\bar\psi c.
\end{align}
The $\mathrm{U}(1)$ part decouples.  The corresponding ghost fields $\tr c$ and
$\tr\bar c$ are free and can be dropped, such that this part of the ST
identity can be replaced by the $\mathrm{U}(1)$ Ward identity~(\ref{singlet-Ward}).

\subsection{Inductive Renormalization}
\label{sec:AR}

For an inductive proof of renormalizability, we can now assume
that we have found a Lagrangian which generates a renormalized
effective action at order $n$ that satisfies all conditions.  The
renormalized order-$n+1$ effective action does not necessarily have
this property, yet.

We do not repeat the standard derivation and just state that the
$\mathrm{SU}(N)$ ST identity~(\ref{ST}) and gauge-fixing
condition~(\ref{gf}) can be extended to order $n+1$, possibly by
adding non-invariant local counterterms to the Lagrangian.
Clearly, we can handle the gauge-fixing condition~(\ref{gf-phantom})
in analogy to the $\mathrm{SU}(N)$ gauge-fixing condition, so this equation can
also be established at order $n+1$.

The only non-standard part is the phantom
equation~(\ref{phantom-eq-exp}).  We do not know whether the
right-hand side is zero at order $n+1$.  The quantum action
principles~\cite{Lowenstein:1971jk,Lam:1972mb,Clark:1976ym,%
  Breitenlohner:1977hr} tell
us, however, that the equation must take the form
\begin{equation}
\label{phantom-eq-anom}
  \tr\frac{\delta\Gamma^{(n)}}{\delta A^\nu}
  + \frac{\delta\Gamma^{(n)}}{\delta\tilde A^\nu}
  - \frac{2}{g^2}\left(\pd^\mu \tr G_{\mu\nu} -
     N \pd^\mu\tilde G_{\mu\nu}\right)
  + \frac{2}{g^2}\left(\pd_\nu\tr B -  N \pd_\nu\tilde B\right) 
  = X_\nu^{(n+1)},
\end{equation}
where the effective action $\Gamma^{(n)}$ is evaluated from the
order-$n$ Lagrangian, including terms in the result up to order $n+1$,
i.\,e., diagrams with $n+1$ loops.  The operator $X_\nu^{(n+1)}$ on the
right-hand side is a local polynomial in fields and derivatives of
dimension three with the same quantum numbers as the left-hand side.

Taking the divergence of~(\ref{phantom-eq-anom}), we obtain
\begin{align}
\label{WI-A+-Anom}
  \pd^\nu\tr\frac{\delta\Gamma^{(n)}}{\delta A^\nu}
  + \pd^\nu\frac{\delta\Gamma^{(n)}}{\delta\tilde A^\nu}
  + \frac{2}{g^2} \left(\pd^2\tr B - N \pd^2\tilde B\right)
  &= \pd^\nu X^{(n+1)}_\nu
\end{align}
This is precisely the Ward identity for the $\mathrm{U}(1)_+$
symmetry~(\ref{WI-A+}) with an anomaly on the right-hand side.
We have to verify that the right-hand side is in fact zero.

It is well known that the only possible obstruction to such a $\mathrm{U}(1)$
Ward identity is the Adler-Bell-Jackiw (ABJ) anomaly~\cite{ABJ},
i.\,e., an operator corresponding to a 
triangle diagram with three external gauge bosons.  In a vector-like
theory such as QCD, this vanishes.  Nevertheless, we may consider a
chiral $\mathrm{SU}(N)$ theory where this is not necessarily the case.

With the hypothesis that the phantom equation~(\ref{phantom-eq-exp})
is valid for the renormalized fields and interactions at order~$n$, we
can apply the change of basis~\eqref{eq:redef} to the
order-$n$ renormalized fields and rewrite this equation as
\begin{align}
\label{phantom-eq-n}
  \frac{\delta\Gamma^{(n)}}{\delta A^{(n)}_{+\nu}}
  - \frac{2N}{g^2}\left(\pd_\mu G^{(n)}_-{}^{\mu\nu}
  - \pd^\nu\tr B^{(n)}_-\right)
  &= 0,
\end{align}
in complete analogy with the tree-level equation~(\ref{phantom-eq}).
Taking the second derivative with respect to matter fields of this relation, we
immediately see that $A^{(n)}_+$ does not interact at all.  In
particular, the interaction $(\bar\psi\fmslash A_+\psi)^{(n)}$ vanishes.
Hence, there is no triangle diagram
involving $A_+$ that can contribute to the ABJ anomaly.  Integrating
this zero, we conclude that the right-hand side of~(\ref{WI-A+-Anom})
vanishes at order~$n+1$.  The $\mathrm{U}(1)_+$ symmetry is preserved.\footnote{Actual
  calculations are carried out in the original basis $(A,\tilde A)$.
  In this basis, the argument implies exact cancellation between
  the integrands of loop graphs involving $A$ and $\tilde A$.  This is
  analogous to the Standard Model, where the absence of anomalies
  is evident in the gauge basis ($W^{\pm 0},B$), but involves a similar
  relation between $\gamma$ and $Z$ interactions in the physical basis
  $W^{\pm},Z,\gamma$.}

Since the right-hand side of~(\ref{WI-A+-Anom}) vanishes, we have
$\pd^\nu X^{(n+1)}_\nu = 0$.  A divergenceless vector must be the
derivative of an antisymmetric tensor, thus
\begin{equation}
  X^{(n+1)}_\nu = \pd^\mu Y^{(n+1)}_{\mu\nu}.
\end{equation}
$Y^{(n+1)}$ is a $\mathrm{SU}(N)$-invariant tensor of dimension two, a local
polynomial of the fields.  The only possibility is
\begin{equation}
  Y^{(n+1)}_{\mu\nu} = a^{(n+1)} \tr{G_{\mu\nu}} + b^{(n+1)}\tilde G_{\mu\nu}
\end{equation}
with constants $a^{(n+1)}$ and $b^{(n+1)}$.  If we now modify the
Lagrangian by the finite counterterms
\begin{equation}
\label{GG-counterterms}
  \Delta\LL = \frac12 a^{(n+1)}\tr G_{\mu\nu} \tr G^{\mu\nu}
              + \frac12 b^{(n+1)}\tilde G_{\mu\nu}\tilde G^{\mu\nu},
\end{equation}
and re-evaluate Eq.~(\ref{WI-A+-Anom}), this term is cancelled.  

In
summary, by adding local counterterms to the Lagrangian, we are able
to satisfy all required conditions to order $n+1$.  In particular, we
establish the vanishing of all nontrivial interactions of $A_+$, i.e.,
the exact cancellation of the singlet and phantom interactions.

We note that the cancellation does not apply to the
orthogonal combination~$A_-$, so in the presence of chiral matter, an
anomaly operator involving $A_-$ is allowed.  However, as argued in
Sec.~\ref{sec:cancellations}, as long as $A_+$ does not interact, an
operator which involves $A_-$, even if it formally breaks the $\mathrm{U}(1)_-$
gauge symmetry, has no observable effect and can be ignored.

This completes the inductive proof of renormalizability.  Unitarity of
the $S$-matrix is then established to all orders by the ST
identity and by the phantom equation, as argued in
Sec.~\ref{sec:cancellations}.

In practice, adding those counterterms is necessary and natural.  In
the usual $\overline{\textrm{MS}}$ renormalization scheme, the singlet
gluon propagator will receive a renormalization different from the octet
gluon propagator, because only the octet has self-couplings.  In order
to keep the color-flow scheme simple, we require finite counterterms
that restore the equality of the propagator residues, and thus of the
$\mathrm{SU}(N)$ and $\mathrm{U}(1)$ couplings, at the next order.
Then, we also have to renormalize the phantom propagator, i.e., the
$\mathrm{U}(1)'$ coupling, to the very same value.  This renormalization
coincides with~(\ref{GG-counterterms}).  We can do this freely since
the values of the $\mathrm{U}(1)$ and $\mathrm{U}(1)'$ couplings do
not enter any observable quantity.

\section{Applications: Tree-Level Amplitudes}
\label{sec:trees}

The color-flow approach is particularly useful for the automatic calculation
of tree-level (squared) amplitudes.   In this section, we choose a few
simple examples that show how this works in practice.

\subsection{Algorithm}

For any amplitude, we replace the QCD diagrams with quarks and gluons
by corresponding color-flow diagrams.  These contain in place of each
octet gluon, either a $\mathrm{U}(N)$ gluon (double color line) or a phantom
(no color).  The latter appears only in places where the gluon directly
connects two fermion (i.\,e., single-color) lines.  Each vertex gets a
factor $1/\sqrt2$ due to the different normalization of the QCD
coupling.  After squaring the amplitude, the color lines of each
contributing color-flow diagram are connected to the lines of the
interfering complex-conjugated diagram, both for the incoming and the
outgoing state.

Instead of starting with ordinary QCD diagrams, an implementation may
construct the amplitude diagrammatically from scratch, treating $\mathrm{U}(N)$
gluons and quarks of definite color and phantoms as distinct, ordinary
particle species.  The cross section is computed by squaring diagrams,
including all interference terms.  We recall that $\mathrm{U}(N)$ gluons do not
interfere with phantom gluons.  

Each squared or interference diagram has a color weight $W$ which is 
simply a signed integer power of $N$, given by
\begin{equation}
  W = N^L \left(\frac{-1}{N}\right)^{I+E},
\end{equation}
where
\begin{align*}
  L &= \text{Number of distinct closed color lines}
\\
  I &= \text{Number of internal phantom propagators}
\\
  E &= \text{Number of external phantom particles}
\end{align*}
The $N^L$ factor originates from contracting each string of Kronecker
deltas, which represent color conservation at vertices, to a single
$\delta_i^i=N$.  The $-1/N$ factors in the phantom propagators,
including those across the cuts, result directly from the Lagrangian.  

It is possible to absorb the $(-1/N)^I$ factor for internal lines by
including $-1/N$ in the phantom propagator, as done in the Feynman
rules in Appendix~\ref{sec:rules}.  The other factors are applied
after squaring the amplitude.  Alternatively, we could absorb all
$1/N$ factors, but not their signs, in the phantom-fermion coupling.

In automatic calculations, the color-flow approach has the advantage
that the combinatorics of constructing Feynman graphs is already
implemented, so the proliferation of diagrams does not raise a
bookkeeping problem.  (The algorithm should avoid to compute identical
kinematics twice, however.)  Counting distinct colors is rather simple
and can be done, for tree graphs, by looking at the external state.
In fact, common conventions~\cite{Boos:2001cv}
\emph{require} an event-generating program to classify the external
state in terms of color connections.  On the other hand, computing
color factors algebraically in a $T^a$-$f^{abc}$ basis requires some
additional infrastructure, and the transformation to the
color-connection basis has to be done explicitly.

Algorithms that do not expand an amplitude in diagrams but compute off-shell
wave functions recursively~\cite{Caravaglios:1995cd,Kanaki:2000ey,Moretti:2001zz} benefit even more
from the color-flow approach.  Assigning color factors to diagrams is not
applicable there, so a straightforward implementation of QCD color algebra
would require keeping a color degree of freedom in each off-shell wave
function while constructing the amplitude, and applying a color-matrix
multiplication at each vertex.  By contrast, in the color-flow approach each
color line can be understood as labeling an independent particle species,
hence the algorithm need not know about color at all.  It just has to
distinguish particle species and their respective Feynman rules.

In the following examples, we verify the color-flow result against the
equivalent color-algebra result, where we evaluate the appropriate
trace of $T^a$ and $f^{abc}$ matrices directly.  We are not interested
in the kinematical part of the calculation that involves propagators,
Lorentz factors, and integrations over momenta, so we omit them.  We
also leave out the QCD coupling which is attached to each vertex, and
the color-averaging factor for the initial state.  We just quote color
weights.

\subsection{Quark-Antiquark Scattering}

We are looking at the process ($q\neq q'$)
\begin{equation}
  q \bar q' \to q \bar q'
\end{equation}
in pure QCD, so at tree level there is a single diagram, gluon
exchange.  

To eliminate all open color indices, we square the amplitude and sum
over colors.  In the diagrams, squaring connects the final state
of the amplitude and its complex conjugate; this is indicated by a
vertical dashed line.  Similarly, the initial states of the amplitude
and its complex conjugate, the open lines at the left and
right margins are also understood to be pairwise connected.
\begin{align*}
\parbox{40\unitlength}{%
  \includegraphics{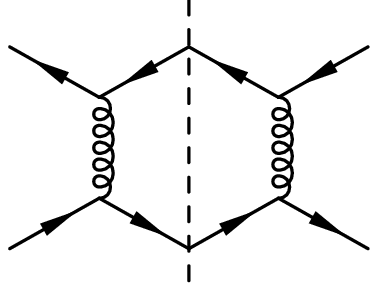}%
}
&\Longleftrightarrow
\quad
\parbox{40\unitlength}{%
  \includegraphics{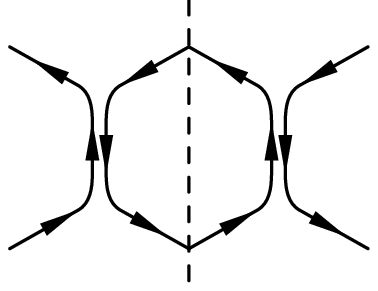}%
}
\\
&\qquad +
\parbox{40\unitlength}{%
  \includegraphics{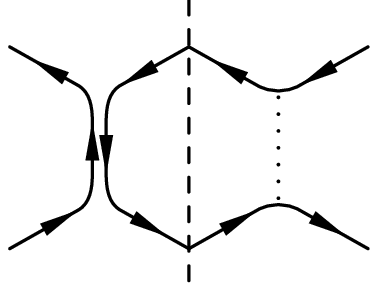}%
}
+
\parbox{40\unitlength}{%
  \includegraphics{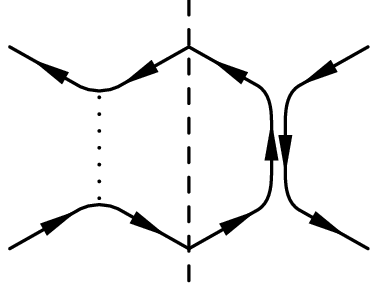}%
}
\\
&\qquad +
\parbox{40\unitlength}{%
  \includegraphics{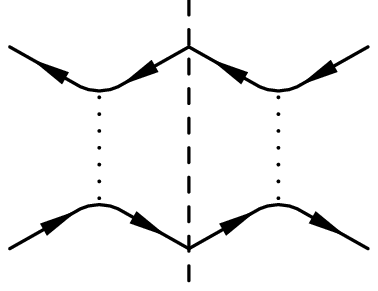}%
}
\end{align*}
For this squared amplitude, standard color algebra yields
\begin{equation}
  W = \tr[T^a T^b]\, \tr[T^a T^b] = \frac14 \delta^{ab}\delta^{ab}
  = \frac14\left(N^2 - 1\right)
\end{equation}
as its color weight.

Color-flow QCD gives four diagrams for the squared amplitude, which
differ only in the color factor.  We recall that there is a factor $N$
for each closed color line, and a factor $-1/N$ for each phantom
propagator.  Including $1/\sqrt2$ for each vertex, we obtain
\begin{align}
  W = \frac14\left(N^2 + N\left(\frac{-1}{N}\right)
     + N\left(\frac{-1}{N}\right) +
     N^2\left(\frac{-1}{N}\right)^2\right)
  &=
  \frac14\left(N^2 - 1\right).
\end{align}
We observe that there is some redundancy in this case which could be
eliminated before computing the result.  In any case, we can read off
the color factor directly without using color algebra.

\subsection{Four-Jet Production in $e^+e^-$}

Let us now consider
\begin{equation}
  e^+e^-\to q\bar q gg,
\end{equation}
where we can ignore the colorless initial state when drawing
color-flow diagrams.  For the amplitude, we have three QCD diagram
structures which decompose into ten color-flow diagrams:
\begin{align}
\parbox{10\unitlength}{%
  \includegraphics{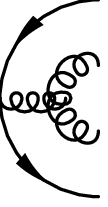}%
}  &\quad\Longleftrightarrow\quad
 \parbox{10\unitlength}{%
   \includegraphics{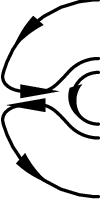}%
}\quad - \quad
 \parbox{10\unitlength}{%
   \includegraphics{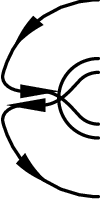}%
}\\[\baselineskip]
\parbox{10\unitlength}{%
  \includegraphics{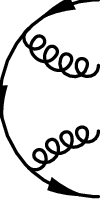}%
}&\quad\Longleftrightarrow\quad
\parbox{10\unitlength}{%
  \includegraphics{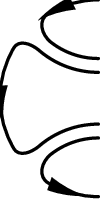}%
}\quad-\quad
\parbox{10\unitlength}{%
  \includegraphics{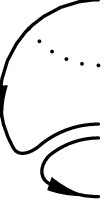}%
}\quad-\quad
\parbox{10\unitlength}{%
  \includegraphics{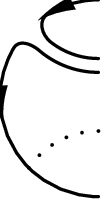}%
}\quad+\quad
\parbox{10\unitlength}{%
  \includegraphics{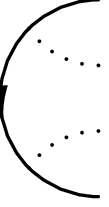}%
}\\[\baselineskip]
\parbox{10\unitlength}{%
  \includegraphics{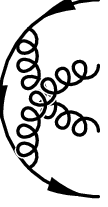}%
}  &\quad\Longleftrightarrow\quad
\parbox{10\unitlength}{%
  \includegraphics{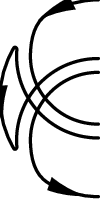}%
}\quad-\quad
\parbox{10\unitlength}{%
  \includegraphics{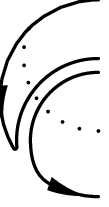}%
}\quad-\quad
\parbox{10\unitlength}{%
  \includegraphics{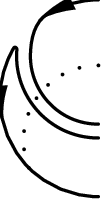}%
}\quad+\quad
\parbox{10\unitlength}{%
  \includegraphics{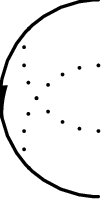}%
}\end{align}
We now look at the squared color-flow diagrams.  Symmetries between
diagrams simplify the calculation.  Summing over all diagrams with
common kinematics, we recover the QCD color weights as linear
combinations of powers of $N$.  We show three terms, the remaining
ones are equivalent regarding their color flow:
\begin{align}
\parbox{30\unitlength}{%
  \includegraphics{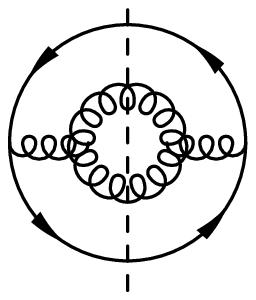}%
}&=
2\times
\parbox{30\unitlength}{%
  \includegraphics{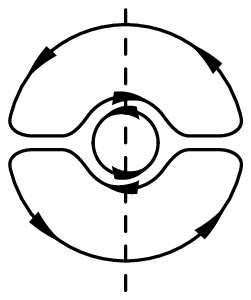}%
}-2\times
\parbox{30\unitlength}{%
  \includegraphics{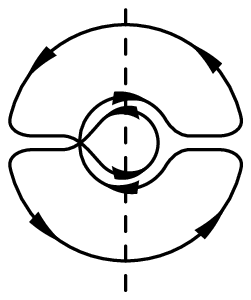}%
}\nonumber\\
&= \frac14\left(2N^3 - 2N\right) = \frac{N}{2}(N^2-1)
\\
\parbox{30\unitlength}{%
  \includegraphics{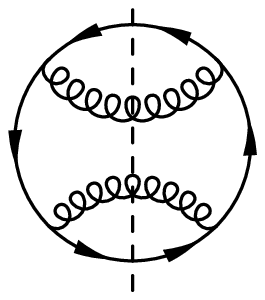}%
}&=
\parbox{30\unitlength}{%
  \includegraphics{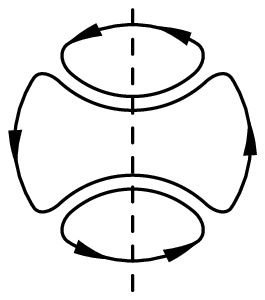}%
}+ 2\times
\parbox{30\unitlength}{%
  \includegraphics{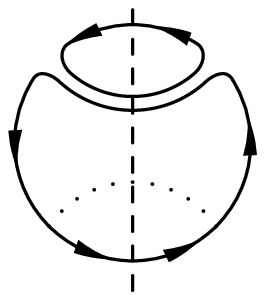}%
}+
\parbox{30\unitlength}{%
  \includegraphics{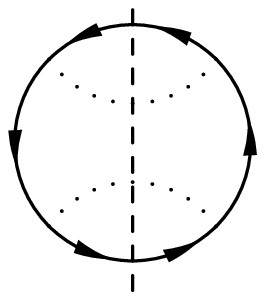}%
}\nonumber\\
&=
\frac14\left(N^3 + 2N^2\left(\frac{-1}{N}\right) 
  + N\left(\frac{-1}{N}\right)^2\right)
= \frac{N}{4}\left(N - \frac{1}{N}\right)^2
\\
\parbox{30\unitlength}{%
  \includegraphics{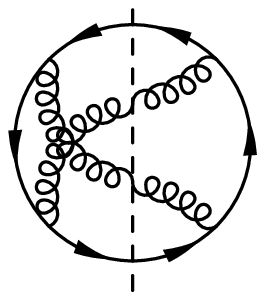}%
}&=
\parbox{30\unitlength}{%
  \includegraphics{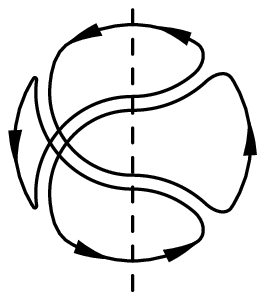}%
}+ 2\times
\parbox{30\unitlength}{%
  \includegraphics{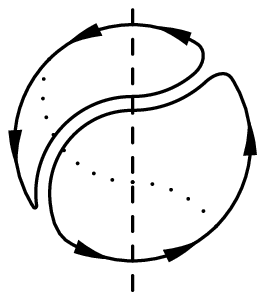}%
}+
\parbox{30\unitlength}{%
  \includegraphics{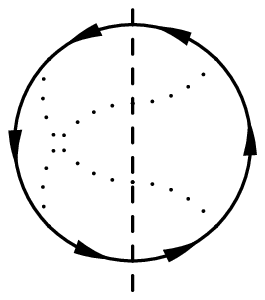}%
}\nonumber\\
&=
\frac14\left(N + 2N^2\left(\frac{-1}{N}\right) 
  + N\left(\frac{-1}{N}\right)^2\right)
= -\frac14\left(N - \frac{1}{N}\right)
\end{align}
In the color-algebra formalism, we compute the color weights
\begin{align}
  -\tr[T^aT^b]\,f^{adc}f^{bcd} &= \frac{N}{2}\left(N^2-1\right),
\\
  \tr[T^aT^aT^bT^b] &= \frac{N}{4}\left(N - \frac1N\right)^2,
\\
  \tr[T^aT^bT^aT^b] &= -\frac{1}{4}\left(N - \frac1N\right),
\end{align}
which agree with the color-flow results, as required.

\section{Applications: Effective Interactions From Loops}
\label{sec:loops}

At tree level, the color-flow method expresses all amplitudes in terms
of Feynman rules that exclusively contain $\mathrm{U}(N)$ gluons and phantom
gluons.  This allows us to compute the color factor of any color-flow
diagram in the expansion of a squared tree-level amplitude by merely
counting the number of distinct \emph{external} color-flow lines, as
detailed in the previous section.  The factors for the internal phantom
propagators are accounted for in the normalization of the propagators.

Once loop diagrams are involved, this is no longer true.  Obviously,
there may be closed color loops not attached to external lines.  As a
further complication, there is a particular class of loop
(sub-)diagrams where singlet gluons make their appearance,
distinguished from the two classes of $\mathrm{U}(N)$ and phantom gluons that
we encounter at tree level.  Fortunately, this only slightly
complicates the algorithm, and computing color factors remains
straightforward.

We are particularly interested in loop amplitudes that can be inserted
as effective vertices in tree-level diagrams.  We absorb the color sum
over internal closed color lines in the corresponding vertex factor.
As a result, we again can deduce the remaining overall color factor of
any amplitude merely looking at the external lines.

An important example is the coupling of Higgs and electroweak bosons
to gluons, which occurs first at one-loop level.  As in the previous
section, we are interested only in color flow, so we ignore all
kinematical and coupling factors.

\subsection{$gg\to H$}

Let us consider colorless particles (in particular, the Higgs boson)
coupled to a gluon pair via a fermion (more generically, matter) loop.
From the color flows in the triangle diagram

\begin{multline}
\parbox{25\unitlength}{%
  \includegraphics{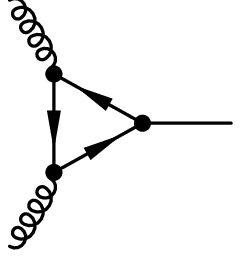}%
} \Longleftrightarrow\parbox{25\unitlength}{%
  \includegraphics{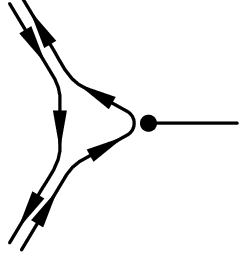}%
}\parbox{25\unitlength}{%
  \includegraphics{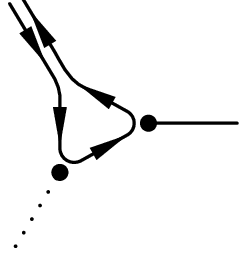}%
}\parbox{25\unitlength}{%
  \includegraphics{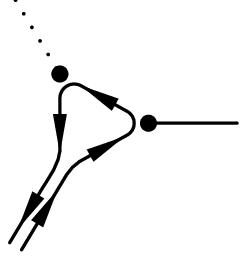}%
}\parbox{25\unitlength}{%
  \includegraphics{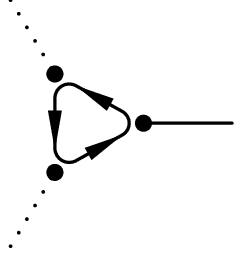}%
}\end{multline}

\vspace{\baselineskip}\noindent
we can derive a set of equivalent ``effective'' Feynman rules

\begin{multline}
\parbox{25\unitlength}{%
  \includegraphics{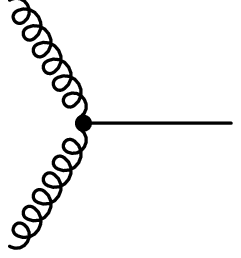}%
} \Longleftrightarrow\parbox{25\unitlength}{%
  \includegraphics{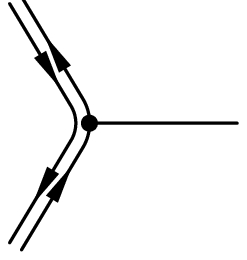}%
}\parbox{25\unitlength}{%
  \includegraphics{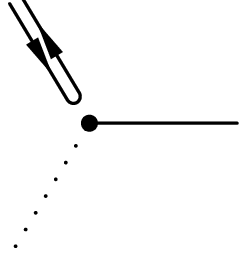}%
}\parbox{25\unitlength}{%
  \includegraphics{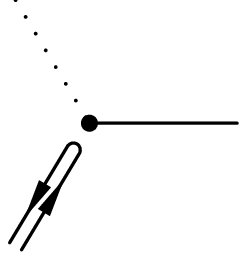}%
}\parbox{25\unitlength}{%
  \includegraphics{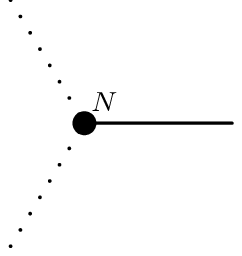}%
}\end{multline}

\vspace{\baselineskip}\noindent
Thus, in the resulting one-loop effective action there appear vertices
that explicitly couple phantom gluons to singlet gluons (and to
themselves).  In the effective action of pure QCD, this is not the
case.  Graphically, we identify the singlet, the projection of a $\mathrm{U}(N)$
gluon, as a color line being reflected at the vertex.  Note that there
is no Feynman rule

\vspace{\baselineskip}
\begin{equation}
\parbox{25\unitlength}{%
  \includegraphics{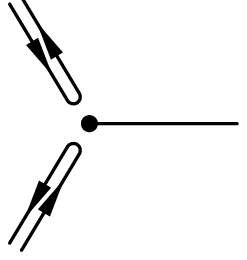}%
}\end{equation}

\vspace{\baselineskip}\noindent
because the singlet-singlet interaction is incorporated in the vertex
where the color lines pass through.

These new Feynman rules for the effective vertex yield the correct
result for $H\to gg$ and, equivalently, $gg\to H$.  Squaring the
amplitude, we get the set of squared diagrams (external Higgs lines
not drawn):
\begin{multline}
\parbox{35\unitlength}{%
  \includegraphics{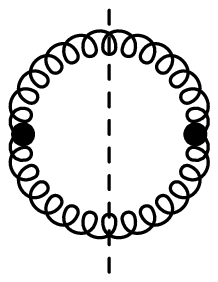}%
} =\parbox{35\unitlength}{%
  \includegraphics{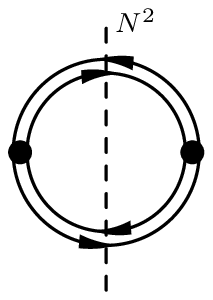}%
} \\ +\parbox{35\unitlength}{%
  \includegraphics{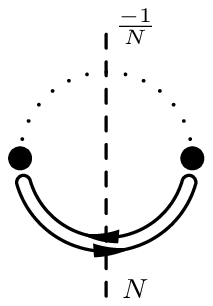}%
} +\parbox{35\unitlength}{%
  \includegraphics{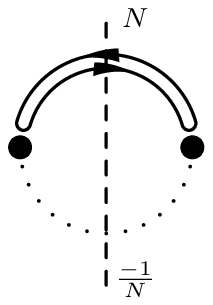}%
} +\parbox{35\unitlength}{%
  \includegraphics{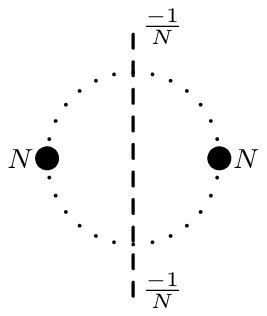}%
} \\ =  N^2 - \frac{N}{N} - \frac{N}{N} + \frac{N^2}{N^2} = N^2 - 1
\end{multline}
Here, there are no interferences.  To get the correct factors, we only
have to remember that each phantom gluon comes with a factor $-1/N$,
while the singlet gluon carries color, which yields a factor $N$ when
summed over.  As in the tree-level case, we just have to count color
lines crossing the cut to obtain the color weights of the squared
diagrams.

\subsection{$gg\to HH$}

Things become interesting when there can be two loop-induced effective vertex
insertions, as in the process $gg\to HH$.  (The irreducible effective
$ggHH$ vertex has the same color structure as $ggH$, so we do not
consider it here.)  In this case, singlet gluons interfere with
$\mathrm{U}(N)$ gluons, projecting out the singlet part on the other
side of the cut:
\begin{multline}
\parbox{35\unitlength}{%
  \includegraphics{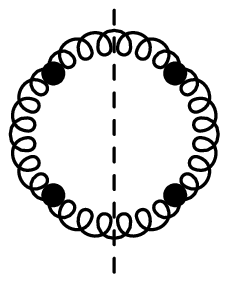}%
} =\parbox{35\unitlength}{%
  \includegraphics{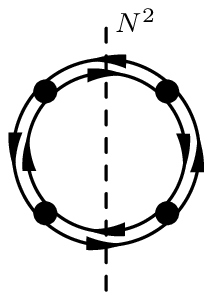}%
} \\ +\parbox{35\unitlength}{%
  \includegraphics{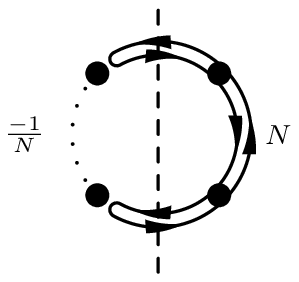}%
} +\parbox{35\unitlength}{%
  \includegraphics{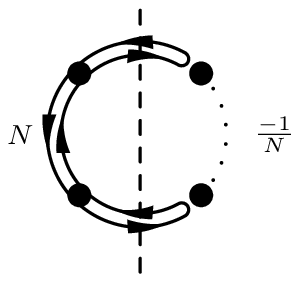}%
} +\parbox{35\unitlength}{%
  \includegraphics{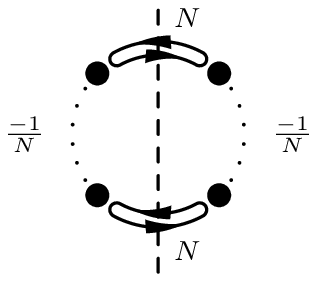}%
} \\ +\parbox{35\unitlength}{%
  \includegraphics{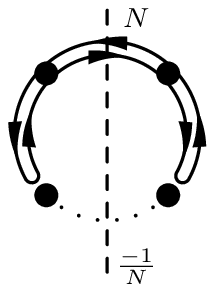}%
} +\parbox{35\unitlength}{%
  \includegraphics{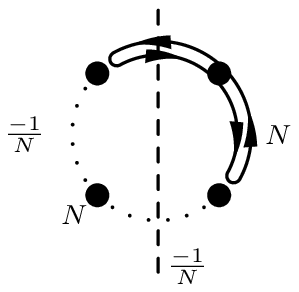}%
} +\parbox{35\unitlength}{%
  \includegraphics{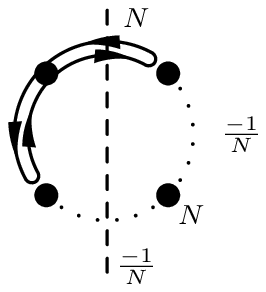}%
} +\parbox{35\unitlength}{%
  \includegraphics{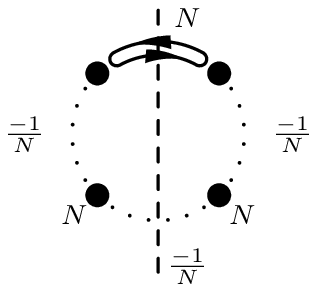}%
} \\ +\parbox{35\unitlength}{%
  \includegraphics{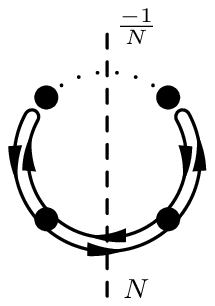}%
} +\parbox{35\unitlength}{%
  \includegraphics{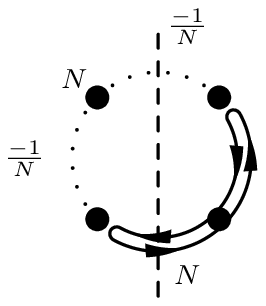}%
} +\parbox{35\unitlength}{%
  \includegraphics{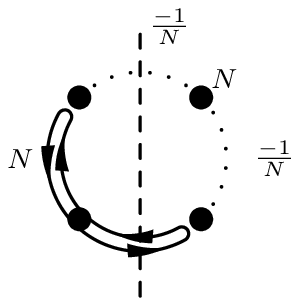}%
} +\parbox{35\unitlength}{%
  \includegraphics{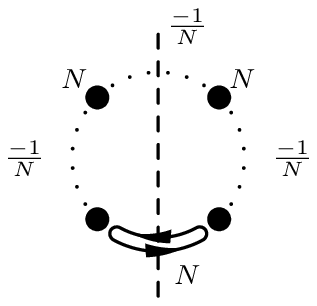}%
} \\ +\parbox{35\unitlength}{%
  \includegraphics{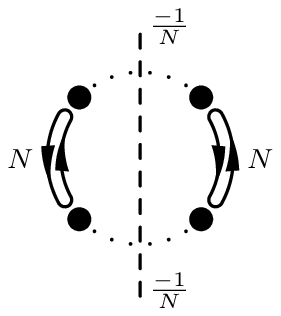}%
} +\parbox{35\unitlength}{%
  \includegraphics{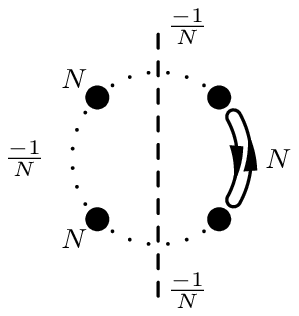}%
} +\parbox{35\unitlength}{%
  \includegraphics{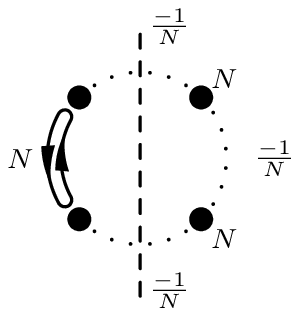}%
} +\parbox{35\unitlength}{%
  \includegraphics{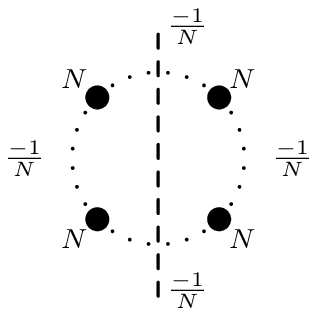}%
} \\ =     N^2
   - 2 \frac{N}{N} +   \frac{N^2}{N^2}
   + 2 \left(
   -   \frac{N}{N} + 2 \frac{N^2}{N^2} -   \frac{N^3}{N^3}
       \right)
                       +   \frac{N^2}{N^2} - 2 \frac{N^3}{N^3} + \frac{N^4}{N^4}
  = N^2 - 1
\end{multline}
Again, the correct factor is recovered, albeit many cancellations are
involved.

At this point, we digress somewhat and discuss a possible way to
remove this redundancy before the sum is computed.  Depending on the
method by which the squared amplitude is computed, this may be useful
for improving efficiency.

We would like to apply the direct cancellation between a phantom and a
singlet line, where it appears obvious.  If the amplitude is
completely expanded in terms of diagrams, explicit singlets appear
always attached to one of the effective vertices that involve a
closed color loop, including the loop pertaining to the effective
non-QCD interaction that we are considering here.  

Starting at this point (the vertex on the left
in the figure), we could apply a procedure that looks like
\begin{equation}
\parbox{35\unitlength}{%
  \includegraphics{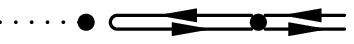}%
}\quad \Longrightarrow\quad
\parbox{35\unitlength}{%
  \includegraphics{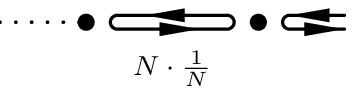}%
}\quad \Longrightarrow\quad
\parbox{35\unitlength}{%
  \includegraphics{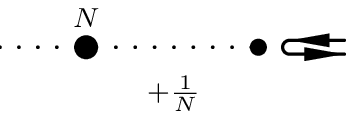}%
}\end{equation}
We can replace the color of an isolated singlet gluon that is attached
to further gluon lines by a closed color loop, if we divide by $1/N$.
We then remove the closed color line from the diagram, attaching the
factor $N$ to the vertex on the left where the singlet originated.
The same factor will be present in the analogous diagram where a
phantom originates from the vertex.  The two contributions now differ
only in sign, and thus cancel.

This algorithm is nonlocal in nature and can be implemented, at face
value, only if the amplitude is expanded in terms of diagrams.  In
automatic computational programs that construct the amplitude
recursively, without expanding Feynman diagrams, the cancellation
procedure would have to be implemented as part of the recursive
calculation.

In any case, if we apply this argument to the $gg\to HH$ amplitude
before squaring, we obtain the much simpler result
\begin{align}
\parbox{20\unitlength}{%
  \includegraphics{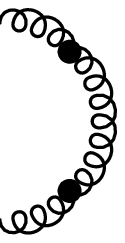}%
}&=
\parbox{20\unitlength}{%
  \includegraphics{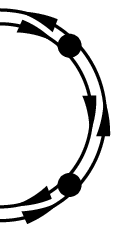}%
}+
\parbox{20\unitlength}{%
  \includegraphics{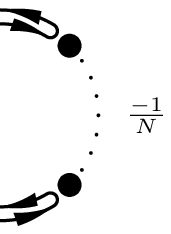}%
}\end{align}
When we compute the square, we have to take into account that singlets, being
contained in the $\mathrm{U}(N)$ gluon matrix, interfere with $\mathrm{U}(N)$
gluons.
This interference acts as a projection
operator which turns a $\mathrm{U}(N)$ gluon into a singlet, on the other side
of the cut:
\begin{align}
\parbox{35\unitlength}{%
  \includegraphics{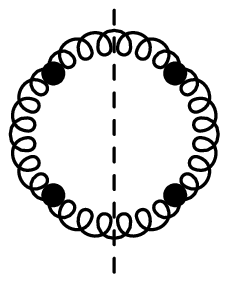}%
} &=
\parbox{35\unitlength}{%
  \includegraphics{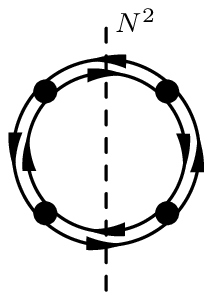}%
} \nonumber\\
&\quad +
\parbox{35\unitlength}{%
  \includegraphics{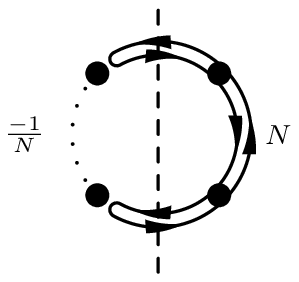}%
} +\parbox{35\unitlength}{%
  \includegraphics{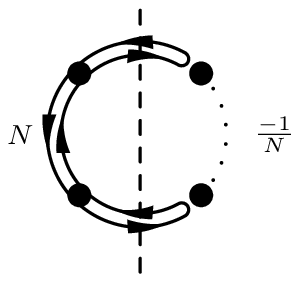}%
} +\parbox{35\unitlength}{%
  \includegraphics{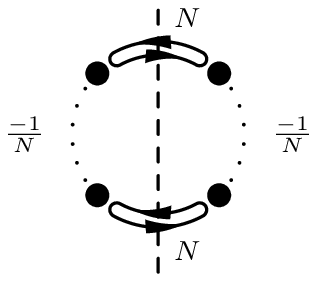}%
}\nonumber\\
 &=
     N^2
   - 2 \frac{N}{N} +   \frac{N^2}{N^2}
 = N^2 - 1
\end{align}
Graphically, we can again apply the cancellation procedure and obtain
the final diagrammatic result
\begin{align}
\parbox{35\unitlength}{%
  \includegraphics{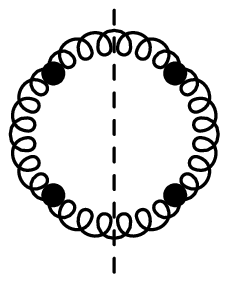}%
} &=
\parbox{35\unitlength}{%
  \includegraphics{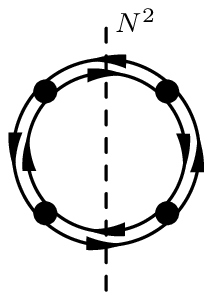}%
}  -
\parbox{35\unitlength}{%
  \includegraphics{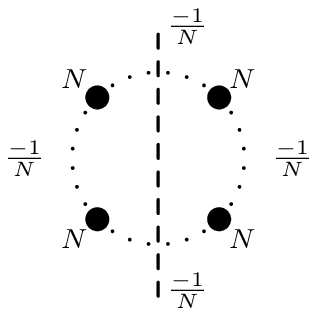}%
}  = N^2 -1
\end{align}
There is no redundancy left.  Note that, effectively, the cancellation
procedure has switched the sign of the phantom-loop graph, so we would
have obtained this result if we just had included this graph with
switched sign, but no graphs that explicitly involve singlets.  This
observation might be generalized and incorporated into the algorithm.


\subsection{$H\to ggg$}

The effective one-loop vertex $H\to ggg$ has two color structures,
$f^{abc}$ and $d^{abc}$, which originate from the difference and sum
of the possible loop orientations, respectively.  In the color-flow
basis, the Feynman rules are
\begin{align}
\parbox{25\unitlength}{%
  \includegraphics{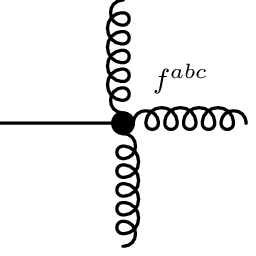}%
}\qquad
&\Longleftrightarrow
\qquad
\parbox{25\unitlength}{%
  \includegraphics{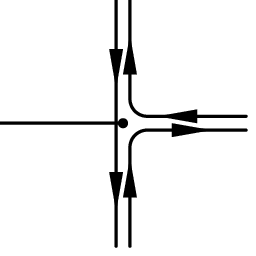}%
}\quad 
-
\quad
\parbox{25\unitlength}{%
  \includegraphics{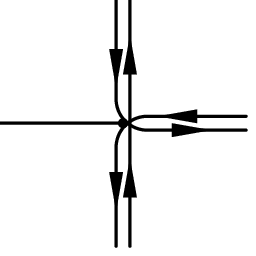}%
}\\ & \nonumber \\
\parbox{25\unitlength}{%
  \includegraphics{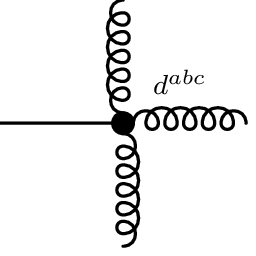}%
}\qquad
&\Longleftrightarrow
\qquad
\parbox{25\unitlength}{%
  \includegraphics{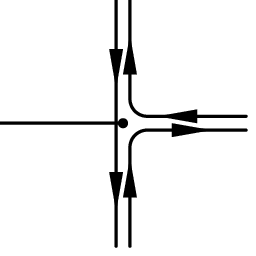}%
}\quad 
+
\quad
\parbox{25\unitlength}{%
  \includegraphics{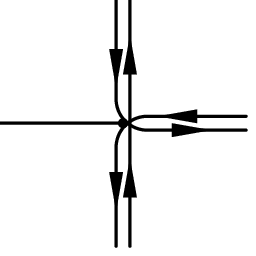}%
}\nonumber\\
&\qquad\qquad
+
\quad
\parbox{25\unitlength}{%
  \includegraphics{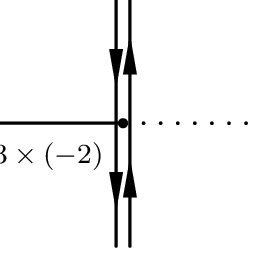}%
}\quad 
+
\quad
\parbox{25\unitlength}{%
  \includegraphics{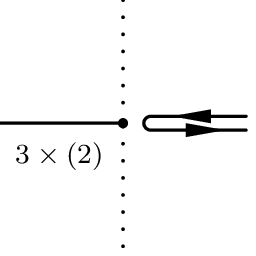}%
}\quad
+
\quad
\parbox{25\unitlength}{%
  \includegraphics{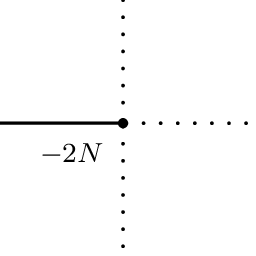}%
}\end{align}
The square of the $f$ term is straightforward:
\begin{align}
  \parbox{25\unitlength}{%
    \includegraphics{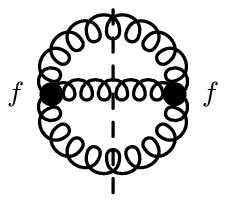}%
}&=
2\times
  \parbox{25\unitlength}{%
    \includegraphics{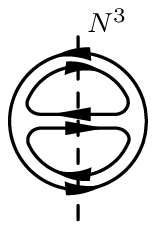}%
}-
2\times
  \parbox{25\unitlength}{%
    \includegraphics{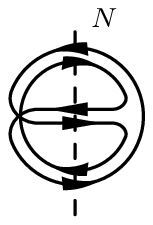}%
}= 2N(N^2-1)
\end{align}
The interference vanishes as expected.  The square of the $d$ term yields
\begin{align}
  \parbox{25\unitlength}{%
    \includegraphics{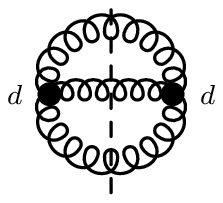}%
}&=
2\times
  \parbox{25\unitlength}{%
    \includegraphics{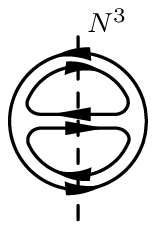}%
}+
2\times
  \parbox{25\unitlength}{%
    \includegraphics{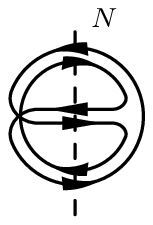}%
}\nonumber\\
&\quad \quad+\quad
  \parbox{25\unitlength}{%
    \includegraphics{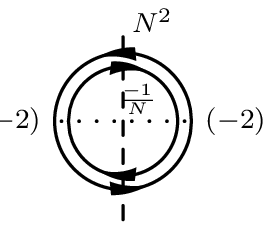}%
}\quad+\quad
  \parbox{25\unitlength}{%
    \includegraphics{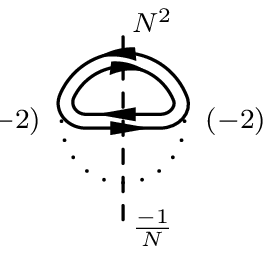}%
}\quad+\quad
  \parbox{25\unitlength}{%
    \includegraphics{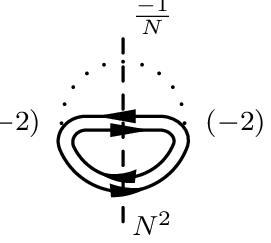}%
}\nonumber\\ \mbox{} & & \nonumber\\ 
&\quad \quad+\quad
  \parbox{25\unitlength}{%
    \includegraphics{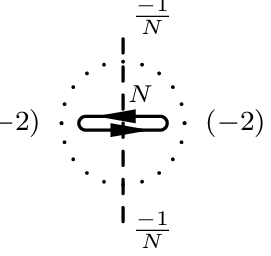}%
}\quad+\quad
  \parbox{25\unitlength}{%
    \includegraphics{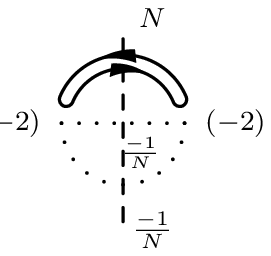}%
}\quad+\quad
  \parbox{25\unitlength}{%
    \includegraphics{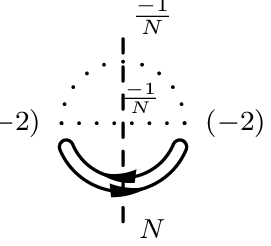}%
}\nonumber\\
&\quad
\quad+\quad\quad
  \parbox{25\unitlength}{%
    \includegraphics{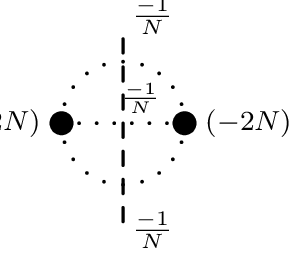}%
}\nonumber\\
&= 2N^3 + 2N - 12N + \frac{12}{N} - \frac{4}{N}
= 2\left(N-\frac{4}{N}\right)(N^2 - 1)
\end{align}

Eliminating the external singlet lines as before would allow us to
combine the last two terms, i.\,e.,
\begin{align}
  2N^3 + 2N - 12N + \frac{8}{N}
= 2\left(N-\frac{4}{N}\right)(N^2 - 1)
\end{align}
which is slightly simpler.

\section{Conclusions}
\label{sec:conclusions}

The expansion of QCD amplitudes in a color-flow basis has been known
as a useful device in various contexts of perturbative and
non-perturbative calculations.  In this paper, we have demonstrated
that it can also be understood as a field theory of its own, a priori
different from, but equivalent to, standard QCD.  We have shown that this field theory
is well defined, renormalizable, unitary, and yields observable
predictions identical to ordinary QCD.

For practical applications, color-flow QCD has advantages
in the context of automatic calculations, where it systematically
generates color-connected amplitudes that can be matched to
parton-shower and hadronization algorithms, avoiding some extra effort
that pertains to color algebra and change of bases.  In particular, it
is useful for algorithms which do not work with a diagrammatic
expansion of the amplitude.

We have implemented color-flow QCD in the O'Mega matrix element
generator, which is the tool for tree-level amplitude generation
contained in the WHIZARD event generator
package~\cite{Kilian:2007gr} together with its parton shower
generator~\cite{Kilian:2011ka}.  The color-flow approach also provides a
convenient way of incorporating exotic color interactions.  If desired, the
methods presented in the present paper can readily be extended to
other exotics, e.\,g., color-decuplet fields.

For higher-order calculations, color-flow QCD can be taken at face
value, provided the renormalization procedure is properly implemented
(cf.\ Sec.~\ref{sec:renormalization}), which is essentially trivial.
Concerning mixed QCD-electroweak processes at loop level, the examples
in Sec.~\ref{sec:loops} show that there are minor technical
complications, which however do not invalidate the procedure.
A comprehensive treatment of NLO calculations would also require a
color-flow exposition of (dipole) subtraction and parton splitting
kernels.  This is not discussed in the present paper, but
straightforward and part of a different publication~\cite{dipoles}.

Finally, we remark that the methods developed in the present paper could be
generalized and applied to other gauge groups, including exceptional groups,
using the completeness relations from~\cite{Cvitanovic:1976am}.

\subsection*{Acknowledgments}

We acknowledge support by the Helmholtz Alliance ``Physics at the Terascale''.
WK has been supported by the BMBF, Contract \#05H09PSE, TO has been supported by
the BMBF, Contract \#05H09WWE. CS has been supported by the Deutsche
Forschungsgemeinschaft through the Research Training Group GRK\,1102
\textit{Physics of Hadron Accelerators}.  WK wants to thank S.~Willenbrock and
T.~Stelzer for illuminating discussions and hospitality at the University of
Urbana-Champaign.

\appendix
\section{Feynman Rules for Color-Flow QCD}
\label{sec:rules}

Here, we list the Feynman rules for $\mathrm{SU}(N)$ gluons interacting with fermions
in the fundamental representation, in the color-flow formalism.  We
shift the
coupling $g/\sqrt{2}$ from the kinetic gluon terms in~(\ref{QCD})
and~(\ref{QCD-phantom}), respectively, to
the vertices by renormalizing gluon and phantom fields accordingly, to
obtain canonically normalized kinetic terms.
Expanding in terms of components, the
color-flow Lagrangian becomes
\begin{align}
\label{QCD-phantom-comp}
  \LL  &= 
    -\frac{1}{4} (G^i_{\;j})_{\mu\nu} (G^j_{\;i})^{\mu\nu}
    +\frac{N}{4}{\tilde G}_{\mu\nu} {\tilde G}^{\mu\nu}
    +\bar\psi_i\left[\ii\fmslash\pd \delta^i_j
           + \frac{g}{\sqrt2}\left({\fmslash A}^i_{\;j} 
             - {\fmslash{\tilde A}}\delta^i_j\right)\right]\psi^j
\nonumber\\
  &\quad
    + 
    B^i_{\;j}(\pd\cdot A^j_{\;i}) - N \tilde B(\pd\cdot \tilde A)
    + \frac{\xi}{2} B^i_{\;j}B^j_{\;i} -  N \frac{\xi}{2}{\tilde B}^2
    + \LL_{\text{ghost}}
\end{align}
where
\begin{align}
\label{QCD-phantom-ghost-comp}
  \LL_{\text{ghost}} &=
    - {\bar c}^i_{\;j}\pd^\mu
    \left(\pd_\mu c^j_{\;i} - \frac{g}{\sqrt2}\ii[A_\mu,c]^j_{\;i}\right).
\end{align}
Note that in the color-flow formalism, the basic coupling
emerges as $g/\sqrt2$.   In the Feynman rules below, Lorentz and
momentum factors are omitted, they retain their usual form.

\vspace{\baselineskip}
\noindent Propagators:
\begin{align}
\parbox{25\unitlength}{%
  \includegraphics{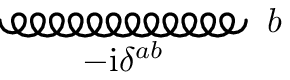}%
}&\qquad\Longleftrightarrow\qquad
\parbox{25\unitlength}{%
  \includegraphics{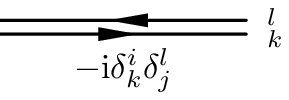}%
}\qquad\qquad
\parbox{25\unitlength}{%
  \includegraphics{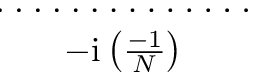}%
}\\
\parbox{25\unitlength}{%
  \includegraphics{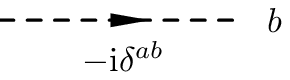}%
}&\qquad\Longleftrightarrow\qquad
\parbox{25\unitlength}{%
  \includegraphics{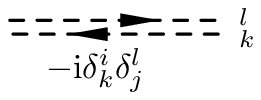}%
}\end{align}
Vertices:

\vspace{\baselineskip}
\begin{align}
\parbox{25\unitlength}{%
  \includegraphics{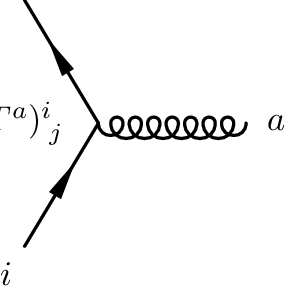}%
} &\qquad\Longleftrightarrow\qquad
\qquad
\parbox{25\unitlength}{%
  \includegraphics{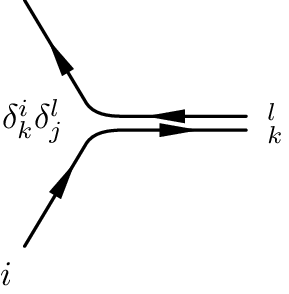}%
}\qquad\qquad
\parbox{25\unitlength}{%
  \includegraphics{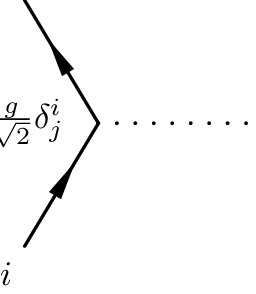}%
} \\[2\baselineskip]
\parbox{25\unitlength}{%
  \includegraphics{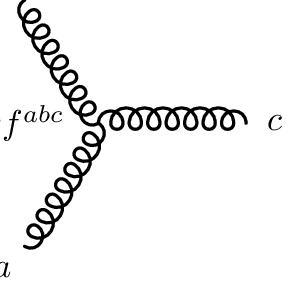}%
} &\qquad\Longleftrightarrow\qquad
\qquad\qquad
\parbox{25\unitlength}{%
  \includegraphics{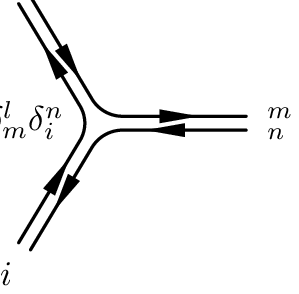}%
} \qquad\qquad\qquad
\parbox{25\unitlength}{%
  \includegraphics{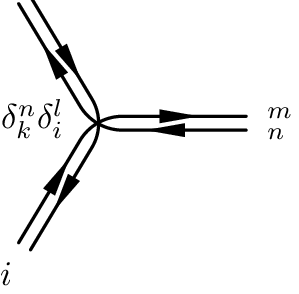}%
} \\[2\baselineskip]
\parbox{25\unitlength}{%
  \includegraphics{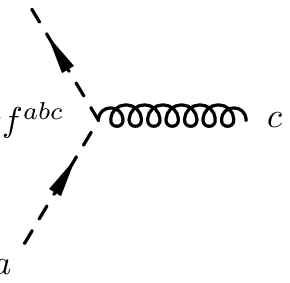}%
} &\qquad\Longleftrightarrow\qquad
\qquad\qquad
\parbox{25\unitlength}{%
  \includegraphics{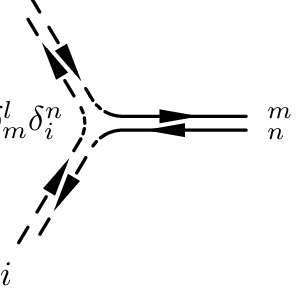}%
} \qquad\qquad\qquad
\parbox{25\unitlength}{%
  \includegraphics{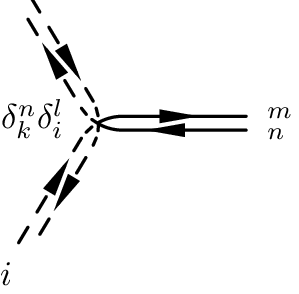}%
} \\[2\baselineskip]
\parbox{25\unitlength}{%
  \includegraphics{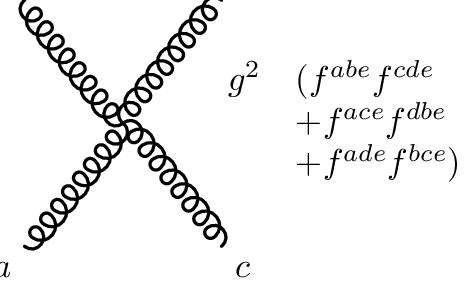}%
} & \\ \notag
& \qquad \Longleftrightarrow
\qquad\qquad
\parbox{25\unitlength}{%
  \includegraphics{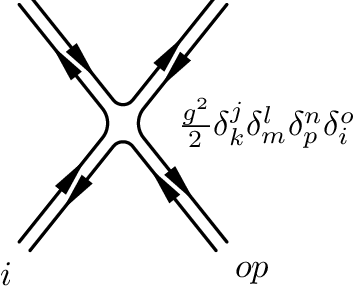}%
} \qquad\qquad
\parbox{25\unitlength}{%
  \includegraphics{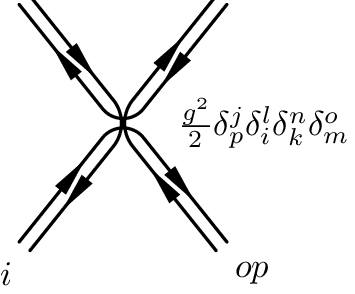}%
} \\[2\baselineskip]  \notag
&
\qquad\qquad\qquad\phantom{\Longleftrightarrow}
\parbox{25\unitlength}{%
  \includegraphics{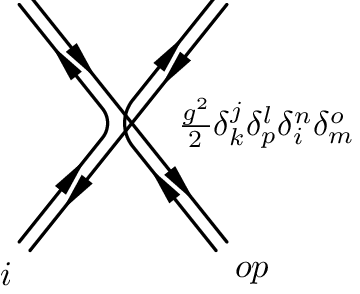}%
} \qquad\qquad
\parbox{25\unitlength}{%
  \includegraphics{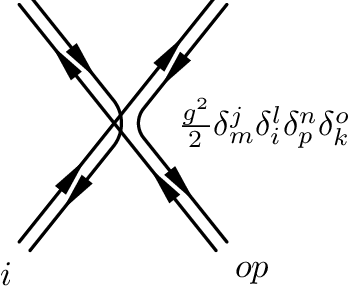}%
} \\[2\baselineskip] \notag
& 
\qquad\qquad\qquad\phantom{\Longleftrightarrow}
\parbox{25\unitlength}{%
  \includegraphics{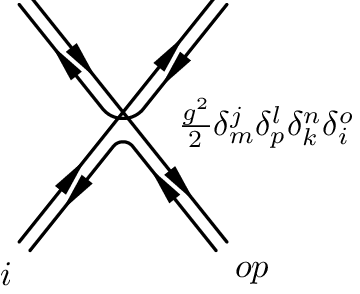}%
} \qquad\qquad
\parbox{25\unitlength}{%
  \includegraphics{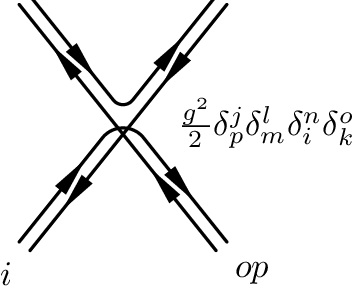}%
} \end{align}
\vspace{2\baselineskip}

\section{Color-Sextet Particles}
\label{sec:exotics}

QCD and the Standard Model contain colored fields only in the
fundamental and in the adjoint representation of $\mathrm{SU}(3)$.
However, various Standard Model extensions that are currently under
discussion provide extra colored fields and interactions.  Automatic
computation programs should be able to deal with 
such extensions, and the color-flow approach provides a
straightforward means to achieve this.  As a particular example, we
describe the color-flow Feynman rules for color-sextet scalar
particles.

Color-sextet particles or, more generally, particles in the symmetric
bi-fundamen\-tal representation of $\mathrm{SU}(N)$ might exist at mass
scales accessible at the 
LHC.  A possible source are extended Higgs (super-)multiplets in
unified gauge theories of the strong and electroweak interactions.

We consider a color-sextet scalar particle denoted as $\sigma$.
An explicit representation is a complex symmetric matrix in color
space with six independent entries $\sigma_{1,2,3}$ and
$\bar\sigma_{1,2,3}$,
\begin{equation}
  \sigma =
  \begin{pmatrix}
    \sigma_1 & \frac{1}{\sqrt2}\bar\sigma_3 & \frac{1}{\sqrt2}\bar\sigma_2
    \\
    \frac{1}{\sqrt2}\bar\sigma_3 & \sigma_2 & \frac{1}{\sqrt2}\bar\sigma_1
    \\
    \frac{1}{\sqrt2}\bar\sigma_2 & \frac{1}{\sqrt2}\bar\sigma_1 & \sigma_3
  \end{pmatrix}
\end{equation}
with components $\sigma_{ij}$, where $\sigma_{ij}=\sigma_{ji}$.  We
also need the antiparticle $\sigma^{ij}$.  This particle couples to
gluons and gluon pairs.  It may also couple to colorless states (e.\,g.,
Higgs), and linearly to quark pairs.  The Lagrangian is
\begin{equation}
  \LL = D_\mu\sigma_{ij}D^\mu\sigma^{ij}
        - m^2 \sigma_{ij}\sigma^{ij}
        - \lambda H\sigma_{ij}\sigma^{ij}
        - g_{q q'} (\sigma^{ij}\bar q_i(q'_j)^c + \text{h.c.})
\end{equation}
The covariant derivative, in color-flow QCD, is given by
\begin{equation}
  D_\mu\sigma 
  = \pd_\mu\sigma - \ii(A_\mu\sigma + \sigma A_\mu^T - 2\tilde A_\mu\sigma)
\end{equation}

We derive the Feynman rules.  The propagator has two components which
symmetrize the color flow:
\begin{equation}
  \parbox{35\unitlength}{%
    \includegraphics{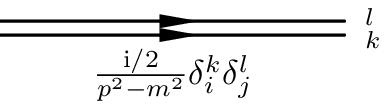}%
}\qquad\qquad
  \parbox{35\unitlength}{%
    \includegraphics{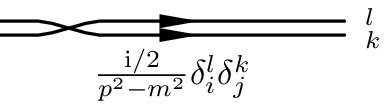}%
}\end{equation}

\vspace{\baselineskip}\noindent
The same symmetrization is needed for external sextet particles.

Gluons interact symmetrically with both color lines.  The single-gluon
interaction splits into three distinct Feynman rules:
\vspace{\baselineskip}
\begin{gather}
  \parbox{35\unitlength}{%
    \includegraphics{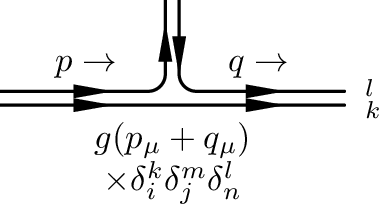}%
}\qquad\qquad
  \parbox{35\unitlength}{%
    \includegraphics{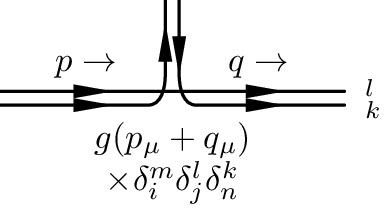}%
}\qquad\qquad
  \parbox{35\unitlength}{%
    \includegraphics{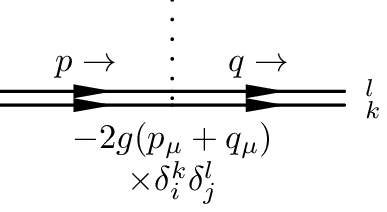}%
}\end{gather}
\vspace{2\baselineskip}
The double-gluon
interaction splits into six distinct Feynman rules:
\begin{gather}
  \parbox{35\unitlength}{%
    \includegraphics{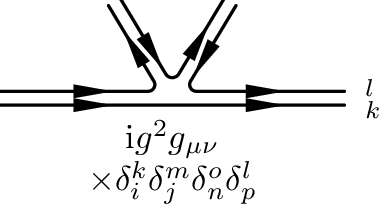}%
} \qquad\qquad
 \parbox{35\unitlength}{%
    \includegraphics{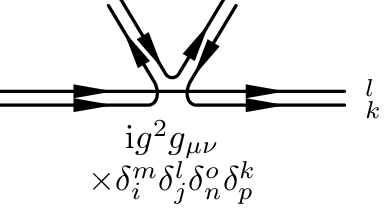}%
}\qquad\qquad
  \parbox{35\unitlength}{%
    \includegraphics{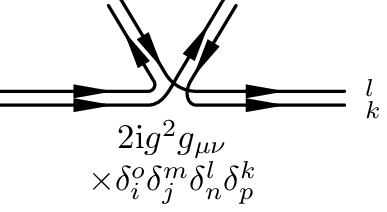}%
}\nonumber\\\nonumber\\\nonumber\\
  \parbox{35\unitlength}{%
    \includegraphics{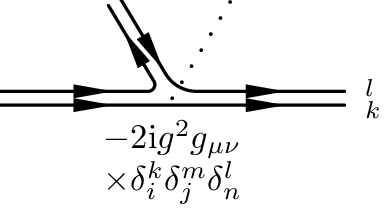}%
}\qquad\qquad
  \parbox{35\unitlength}{%
    \includegraphics{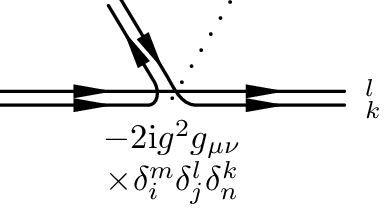}%
}\qquad\qquad
  \parbox{35\unitlength}{%
    \includegraphics{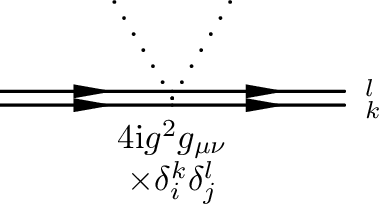}%
}\end{gather}
Couplings to colorless particles are simply
\begin{equation}
  \parbox{35\unitlength}{%
    \includegraphics{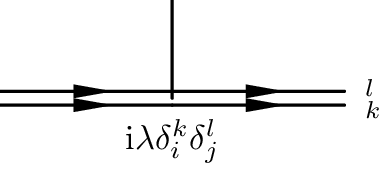}%
}\end{equation}
and the coupling to quark pairs has the Feynman rules
\begin{equation}
  \parbox{35\unitlength}{%
    \includegraphics{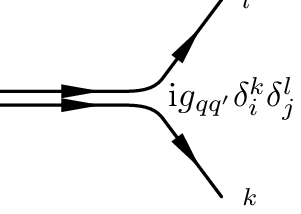}%
}\qquad
  \parbox{35\unitlength}{%
    \includegraphics{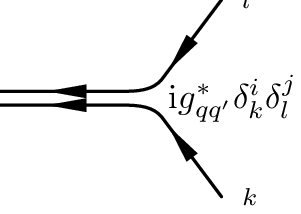}%
}\end{equation}

\end{fmffile}

\baselineskip15pt
\bibliographystyle{JHEP}

\end{document}